\documentclass[twocolumn,twocolappendix]{aastex631}

\usepackage{graphicx}	
\usepackage{amsmath}	
\usepackage{subfigure}
\usepackage{natbib}
\usepackage{color}
\usepackage{tabularx}
\usepackage{longtable}
\usepackage{bm}
\usepackage{threeparttable}
\usepackage{enumerate}
\usepackage[normalem]{ulem}
\usepackage{verbatim}
\usepackage[T1]{fontenc}

\newcommand{\code}[1]{{\texttt{#1}}}
\newcommand{\tess}{{\it TESS}}

\newcommand{\gaia}{{\it Gaia}}

\newcommand{\tar}{{HD 203771}}

\accepted{October 12, 2026}
\submitjournal{the Astronomical Journal}

\graphicspath{{./}{figures/}}

\begin{document}

\title{An Intermediate-orbit Brown Dwarf Confirmed with Astrometry+Radial Velocity Joint Analysis, and the Eccentricity Distribution of Brown Dwarfs Across Different Locations}

\correspondingauthor{Tianjun Gan}
\email{tianjungan@gmail.com}

\author[0009-0003-3412-4773]{Lei Ren}
\affil{Department of Astronomy, Westlake University, Hangzhou 310030, Zhejiang Province, People’s Republic of China}

\author[0000-0002-4503-9705]{Tianjun~Gan}
\affil{Instituto de Astrof\'{i}sica de Canarias (IAC), E-38205 La Laguna, Tenerife, Spain}
\affil{Departamento de Astrof\'isica, Universidad de La Laguna (ULL), E-38206 La Laguna, Tenerife, Spain}

\author[0009-0000-9341-3442]{Yifan Xuan}
\affil{State Key Laboratory of Dark Matter Physics, Tsung-Dao Lee Institute \& School of Physics and Astronomy, Shanghai Jiao Tong University, Shanghai 201210, China}

\author[0000-0001-8317-2788]{Shude Mao}
\affil{Department of Astronomy, School of Science, Westlake University, Hangzhou 310030, Zhejiang Province, People’s Republic of China}

\author[0000-0002-9676-3891]{Shigeru Ida}
\affil{Department of Astronomy, Westlake University, Hangzhou 310030, Zhejiang Province, People’s Republic of China}

\author[0000-0001-6753-4611]{Guang-yao Xiao}
\affil{State Key Laboratory of Dark Matter Physics, Tsung-Dao Lee Institute \& School of Physics and Astronomy, Shanghai Jiao Tong University, Shanghai 201210, China}

\author[0000-0001-6039-0555]{Fabo Feng}
\affil{State Key Laboratory of Dark Matter Physics, Tsung-Dao Lee Institute \& School of Physics and Astronomy, Shanghai Jiao Tong University, Shanghai 201210, China}

\author[0000-0003-0987-1593]{Enric Pall\'{e}} 
\affil{Instituto de Astrof\'{i}sica de Canarias (IAC), E-38205 La Laguna, Tenerife, Spain}
\affil{Departamento de Astrof\'isica, Universidad de La Laguna (ULL), E-38206 La Laguna, Tenerife, Spain}



\begin{abstract}
Intermediate-orbit ($\sim 1$ AU) brown dwarfs with absolute mass measurements are essential for understanding the evolution of substellar objects. Although such systems remain rare, their number is now being enlarged thanks to astrometry. Here, we present the absolute mass measurements for \tar B, a massive brown dwarf orbiting a K dwarf ($M_\ast=0.79\pm0.08\ M_\odot$) with a period of 1140 days on an eccentric ($e\approx0.5$) and inclined ($i\approx 130^{\circ}$) orbit. Through joint analysis of (1) radial velocities and Gaia two-body solution, as well as (2) radial velocities, Hipparcos Intermediate Astrometric Data and two-epoch Gaia astrometry, we obtain true companion masses of $74.20^{+3.46}_{-3.64} M_{\mathrm{J}}$ and $69.03^{+4.99}_{-4.95} M_{\mathrm{J}}$, respectively, agreeing with each other within $1\sigma$. Taking a sample of 159 brown dwarfs with absolute mass measurements from different detection techniques, we investigate their eccentricity distributions across different semi-major axis ranges by fitting a Beta distribution. We find that the peak of eccentricity distribution tends to gradually shift toward lower values from wide orbits ($a>20$~AU) to intermediate ($1<a<20$~AU) and inner ($0.1<a<1$~AU) regions. Such trend is plausibly due to disk friction during orbital evolution. Together with obliquity studies showing that brown dwarfs are mostly aligned, our result suggests that brown dwarfs may migrate inward through quiescent coplanar disk-driven processes. 

\end{abstract}



\section{Introduction} \label{sec:intro}

As intermediate mass objects between the deuterium-burning threshold $13\ M_J$ \citep{Burgasser2003,Spiegel2011} and hydrogen fusion $80\ M_J$ \citep{Laughlin1997}, many bound brown dwarfs have been detected over the past decade using a variety of techniques. However, their formation mechanism, either through the bottom-up core accretion channel similar to planetary systems \citep{Pollack1996,Ida2004II,Mordasini2008}, or via the top-down gravitational instability route that akin to stellar binaries \citep{Boss2000,Helled2014,Kratter2016}, remains under debate. 

Several efforts have been made to explore the demographics of brown dwarfs, aiming to characterize the mass boundary of different formation channels. For example, with a mixed sample of giant planets, brown dwarfs, and low-mass stars orbiting FGK stars, \cite{Cui2026} investigated the occurrence rates of companions across different mass and semi-major axis bins. The authors identified a distinct ``brown dwarf desert'' at around $30\,M_{\mathrm{J}}$ \citep[see also][]{Marcy2000}, suggesting two distinct formation pathways above and below this threshold, and such a paucity feature persists out to 20 AU. They further found that companions above $30\,M_{\mathrm{J}}$ tend to have lower host-star metallicities and higher eccentricities than those below this threshold. Moreover, several studies have investigated the “brown dwarf desert” from the perspective of stellar metallicity distribution (e.g., \citealt{Ma2014,Maldonado2017,Schlaufman2018}). Recently, \cite{Giacalone2026} examined the stellar metallicity distribution of systems with companion masses between 0.5 and 110 $M_{\mathrm{J}}$ and semi-major axes greater than 1~AU, based on the California Legacy Survey (CLS) \citep{Rosenthal2021}. They claimed a similar transition at about $27\,M_{\mathrm{J}}$: the lower-mass population preferentially orbit metal-rich hosts, and those higher-mass objects are more likely to appear around stars with solar-like metallicity. Further evidence comes from studies of orbital eccentricity, which encodes the information of the formation and evolution history of brown dwarfs. For example, \cite{Ma2014} found that brown dwarfs with masses above $42.5\,M_{\mathrm{J}}$ exhibit eccentricities resembling those of stellar binaries, whereas less massive ones follow a distribution approximate to that of giant planets. More recently, \cite{Bowler2020} reported that wide-orbit (5-100 AU) brown dwarfs have a broad eccentricity distribution between 0.6 and 0.9 \citep[see also][]{Doo2023,Nagpal2023}, which differs from cold Jupiters in the same semi-major axis range, with eccentricity distribution centering at about 0.1. In contrast, intermediate-orbit (0.1-1.5 AU) brown dwarfs and warm Jupiters tend to have analogous eccentricity distributions \citep{Gan2025}, though \cite{Subjak2026} argued that the sample size of brown dwarfs is limited and the result is dependent on the choice of period cut \citep[see also][]{Ferreira2026}. Recent work from \cite{Blunt2026} examined the CLS sample over $0.1$-$4.5$ AU semi major axis range and found an eccentricity distribution peaking at $e\simeq0.3$ among super-Jovian companions ($3.1$–$18.9\,M_{\mathrm{J}}$), while a similar peak does not show up for the brown dwarf sample ($18.9$–$94.4\,M_{\mathrm{J}}$). \cite{Gilbert2026} looked into the eccentricity distribution of brown dwarfs in the CLS sample ($1$-$10$ AU) and found the mean eccentricity fall around 0.5, without any peaks. Nevertheless, taken together, these results hint that brown dwarfs and giant planets probably form in the outer regions through different channels, and subsequently migrate inwards under similar orbital evolution processes. 

However, many of the demographic works of brown dwarfs relied on systems detected by radial velocity (RV) surveys, which only provide minimum mass constraints. Astrometric mass refinements from \cite{Unger2023} using Gaia DR3 astrometry have revealed that some RV-detected brown dwarfs are, in fact, low-mass M stars with absolute masses exceeding $80\,M_{\mathrm{J}}$. Therefore, some demographic results might be biased to stellar binaries. 
Ideally, one requires a sample of brown dwarfs with their true mass determined for statistics \citep[e.g.,][]{Gan2025,Vowell2025}. The forthcoming Gaia DR4 time-series astrometry is expected to yield a large amount of cold Jupiters with absolute mass measurements \citep{Perryman2014,Lammers2026}, and similarly the population of intermediate-orbit brown dwarfs will also grow substantially, given that the astrometric signature scales with the mass ratio of the system. Although Gaia DR4 will not be available until the end of 2026, astrometric characterization is already feasible via joint analyses of RVs and Hipparcos+Gaia \citep{Perryman1997,vanLeeuwen2007,Gaia2018,Gaia2023} absolute astrometry \citep{Brandt2019,Feng2022,Xiao2023,Sozzetti2023,Piccinini2026}, or RVs and Gaia DR3 two-body orbital solutions \citep{Halbwachs2023,Holl2023,Winn2022,Gan2023gaia,Fitzmaurice2024,Stefansson2025}, based on Gaia observations spanning about 3 years. 

Here, we revisit the \tar\ system, a long-period brown dwarf on a 3-yr orbit orbiting a K1V dwarf initially reported by \cite{Feng2022}, and measure the absolute mass of companion using the two aforementioned methodologies. The rest of this paper is structured as follows. 
In Section~\ref{data}, we described the RV and astrometry data used in this study. Section~\ref{jointfit} presents the stellar characterization and the joint analyses we performed. We investigate the eccentricity distributions of a sample of brown dwarfs with absolute mass measurements across different semi-major axis ranges in Section~\ref{eccentricity}, before we summarize our findings in Section~\ref{conclusion}. 

\section{Observations}\label{data}
\subsection{HARPS Radial Velocities}\label{harps_data}

\begin{figure*}[htb]
    \centering
    \includegraphics[width=0.99\linewidth]{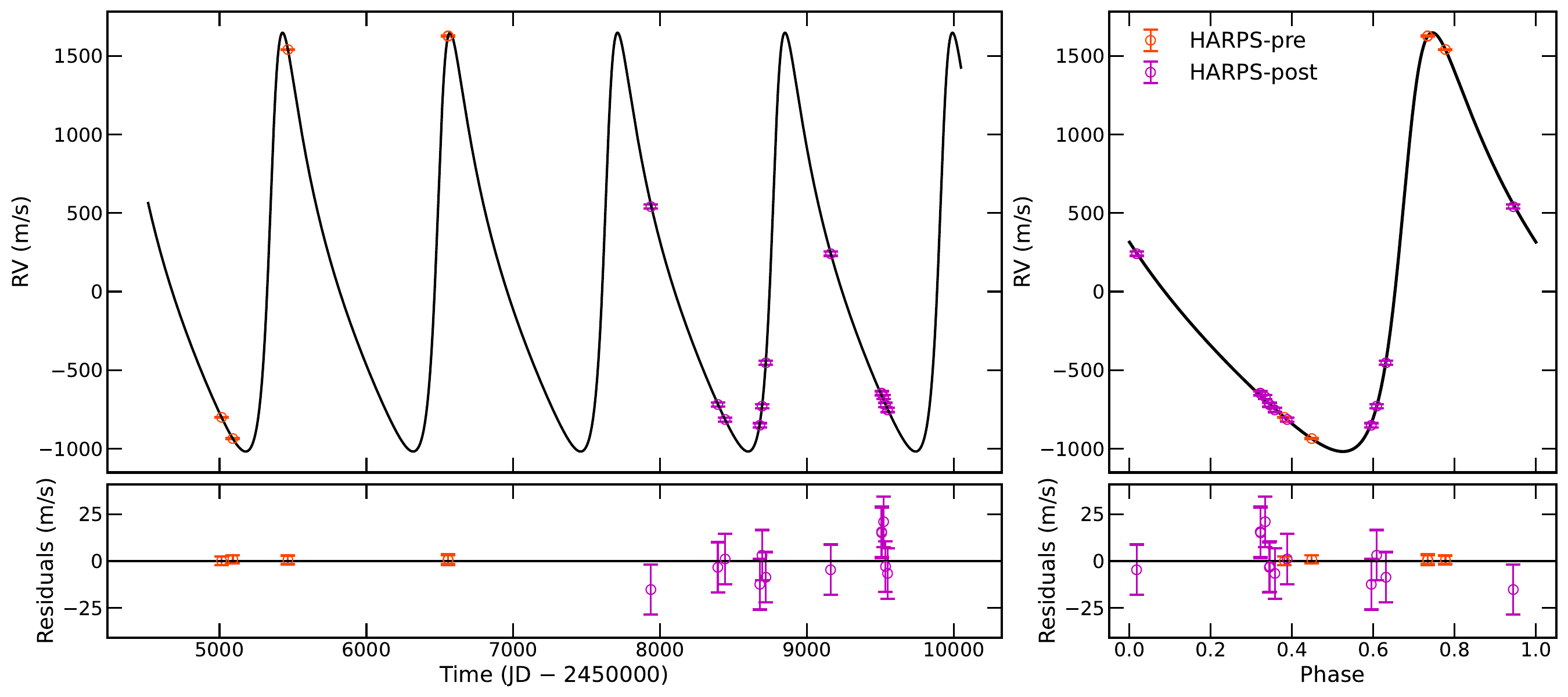}
    \caption{The HARPS RVs of \tar\ after subtracting the best-fit offset $\gamma_{\mathrm{HARPS-pre}}$ and $\gamma_{\mathrm{HARPS-post}}$, along with the Keplerian model shown as a black curve constructed based on the RV + Gaia two-body solution joint analysis. The error bars are the quadrature sum of the individual RV measurement uncertainties and the fitted jitter.}
    \label{fig:rv_fit}
\end{figure*}

We retrieved the publicly available RV data of \tar\ from HARPS \citep{Mayor2003}, installed on the ESO 3.6 m telescope at La Silla Observatory in Chile, from the HARPS-RVBank \citep{Trifonov2020, HARPSRVBank2}. All RVs were homogeneously measured using the \code{SERVAL} package \citep{Zechmeister2018} based on a template matching method. A total of 17 RVs were collected between June 2009 and November 30 2021 in high-accuracy mode with a spectral resolution of $R\approx 115,000$, including 5 and 12 measurements before and after the HARPS optical fiber upgrade happened in June 2015 \citep{Lo2015}. Therefore, we treated the data as if they were obtained from two distinct instruments, and denoted two datasets as HARPS-pre and HARPS-post. We list the HARPS RVs of \tar\ in Table~\ref{RVtable}, and present the data in Figure~\ref{fig:rv_fit}.

\subsection{Astrometry Data}\label{astrometry_data}

We adopted the Gaia DR3 non-single-star (NSS) two-body orbital solution \citep{Halbwachs2023,Holl2023}, which provides the fitted orbital parameters and their associated uncertainties derived from fitting Gaia astrometric time series over the approximately 34-month DR3 observing baseline. This solution is used in the orbital fit described in Section~\ref{two_body_RV_fit}.

We also used the Gaia DR2 and DR3 five-parameter astrometric solutions \citep{Lindegren:2018,Lindegren:2021}, including the source position, parallax, and linear proper motion from the standard astrometric processing. In addition, we retrieved the Hipparcos intermediate astrometric data (IAD) from the new Hipparcos reduction \citep{vanLeeuwen2007}, which provides individual astrometric measurements at specific observation epochs and therefore retain the time-dependent astrometric information. The Gaia DR2/DR3 five-parameter astrometric solutions and Hipparcos IAD are jointly used in the orbital fit described in Section~\ref{joint_fit_2}.

\section{Analysis}\label{jointfit}

\subsection{Stellar Properties}\label{stellar_properties}

We first measured the stellar parameters, including stellar radius $R_\ast$, stellar mass $M_\ast$, effective temperature $T_{\rm eff}$, and iron abundance [Fe/H] of \tar\, based on the HARPS spectroscopic data. We employed the \code{SpecMatch-Emp} package \citep{Yee2017} to match the co-added HARPS spectrum to a high resolution spectroscopic library that consists of about 400 well-characterized stars, constructed with Keck/HIRES observations. We resampled the HARPS spectrum and its uncertainties onto the wavelength grid of the SpecMatch-Emp library using the interpolation routines provided by \code{scipy} \citep{Virtanen2020} before matching. We obtained a stellar radius of $R_\ast=0.79\pm0.10\ R_\odot$, a stellar mass of $M_\ast = 0.79\pm0.08\ M_\odot$, an effective temperature of $T_{\rm eff}=4804\pm110\ K$, and a metallicity of [Fe/H]$=0.14\pm0.09$~dex. 



As a double check, we further performed an analysis of the broadband spectral 
energy distribution (SED) with \code{ARIADNE} \citep{ARIADNE} to determine the stellar properties of \tar\ (Figure~\ref{fig:sed}). We pulled the NUV magnitude from {\it GALEX} \citep{Morrissey2007}, the $B_{\rm T}$, $V_{\rm T}$ magnitudes from {\it Tycho-2} \citep{hog2000}, three {\it Gaia} magnitudes $G$, $G_{\rm BP}, G_{\rm RP}$ from {\it Gaia} DR3 \citep{Gaia2023}, $JHK$ magnitudes from {\it 2MASS} \citep{Cutri2003} as well as $W1$, $W2$ magnitudes from {\it WISE} \citep{wright2010}. We placed normal priors on the stellar distance but uniform priors on effective temperature $T_{\rm eff}$, surface gravity $\log g$ and stellar radius $R_\ast$. We fixed the extinction $A_V$ to zero, due to the proximity of the star. 
We performed the SED fitting using the PHOENIX and Kurucz atmospheric model grids, fitting each model grid separately and combining their posterior distributions through Bayesian model averaging. Nested sampling was performed using the \code{dynesty} package \citep{Speagle2019} to determine the posteriors of each parameter. We adopted standard nested sampling with 500 live points and terminated the sampling when the estimated remaining contribution to the log-evidence reached $\Delta\ln Z=0.5$. Multiple bounding regions and random-walk sampling were used. We obtained a stellar effective temperature of $T_{\rm eff}=4874\pm27$~K, a surface gravity of $\log g=4.52\pm0.09$, a metallicity of [Fe/H]=$0.04\pm0.09$ dex, and a stellar radius of $R_\ast = 0.84\pm0.02\ R_\odot$, which agrees with our findings through spectroscopic analysis above within $1\sigma$. We summarize the final stellar parameters in Table~\ref{stellar_param}.

\begin{table}[h]\scriptsize
    \centering
    \caption{Summary of stellar properties for \tar.}
    \begin{tabular}{lll}
        \hline\hline
        Parameter       &Value       &Ref. \\\hline
        \it{Main identifiers}                    \\
         HIP &105740 & Hipparcos$^{[1]}$\\
         \gaia\ ID            &$6827680034091744896$ &\gaia\ DR3$^{[2]}$\\
         TIC                     &$99647495$   &$\rm TIC\ V8.2^{[3]}$\\
         \it{Equatorial Coordinates} \\
         $\alpha_{\rm J2015.5}$    &~21:25:01.97 &$\rm TIC\ V8.2$\\
         $\delta_{\rm J2015.5}$    &-23:19:10.37   &$\rm TIC\ V8.2$ \\
         \it{Photometric properties}\\
         $\tess$\ (mag)           &$9.044\pm0.006$   &$\rm TIC\ V8.2$  \\
         $B$ (mag) &$11.235\pm0.066$ &Tycho-2$^{[4]}$\\
         $V$ (mag) &$10.022\pm0.035$ &Tycho-2\\
         $G$ (mag)           &$9.6393 \pm 0.0001$   &\gaia\ DR3   \\
         $G_{\rm BP}$ (mag)           &$10.1343 \pm 0.0005$   &\gaia\ DR3   \\
         $G_{\rm RP}$ (mag)           &$8.9879 \pm 0.0002$   &\gaia\ DR3   \\
         $J$\ (mag)                    &$8.204\pm0.024$   &2MASS$^{[5]}$\\
         $H$\ (mag)                    &$7.769\pm0.029$   &2MASS \\
         $K$\ (mag)                    &$7.650\pm0.023$    &2MASS \\
         $W$1 (mag)                   &$7.595\pm0.032$   &WISE$^{[6]}$ \\
         $W$2 (mag)                   &$7.684\pm0.021$   &WISE \\
         $W$3 (mag)                   &$7.636\pm0.019$   &WISE \\
         \it{Astrometric properties}\\
         $\varpi$ (mas)              &$17.73\pm0.19$  &\gaia\ DR3  \\
         $\mu_{\alpha}\ ({\rm mas~yr^{-1}})$     &$119.93\pm0.21$   &\gaia\ DR3   \\
         $\mu_{\delta}\ ({\rm mas~yr^{-1}})$     &$-153.20\pm0.16$   &\gaia\ DR3  \\
         RV\ (km~s$^{-1}$)         &$-49.16\pm 0.21$ &\gaia\ DR3  \\
         \it{Stellar parameters} \\
         Distance (pc)                &$56.41\pm 0.34$  &This work  \\
         $M_{\ast}\ ({\rm M_{\odot}})$ &$0.79\pm 0.08$ &This work$^{[7]}$       \\
         $R_{\ast}\ ({\rm R_{\odot}})$ &$0.79\pm 0.10$ &This work       \\
         $\log g_{\ast}\ ({\rm cgs})$       &$4.52\pm 0.09$  &This work        \\
         $T_{\rm eff}\ ({\rm K})$           &$4804 \pm 110$  &This work       \\
         $\rm [Fe/H]$ (dex)  &$0.14\pm 0.09$ &This work \\
         \hline
    \end{tabular}
    \begin{tablenotes}
    \item[1]  [1]\cite{Perryman1997}; [2]\cite{Gaia2023}; [3]\cite{Stassun2019tic}; [4]\cite{hog2000}; [5]\cite{Cutri2003}; [6]\cite{wright2010}; [7]We report the stellar properties from spectroscopic analysis.
    \end{tablenotes}
    \label{stellar_param}
\end{table}

\subsection{Joint Fit of RVs and Gaia Two-body Orbital Solution} \label{two_body_RV_fit}

Following the methodology proposed in \cite{Winn2022}, we perform a joint fit of RVs together with the Gaia two-body orbital solution of \tar, using the algorithm developed in \cite{Gan2023gaia}. In short, we maximize the Gaia+RV likelihood function through Markov-Chain Monte Carlo (MCMC) sampling. We detail the construction of the joint likelihood function below.

The Gaia two-body solution contains eight best-fit orbital parameters along with a correlation matrix between them \citep{Halbwachs2023,Holl2023}:
\begin{equation}\label{two_body_param}
     \{ A,\ B,\ F,\ G,\ P,\ t_{p},\ e,\ \varpi \},
\end{equation}
where $A,\ B,\ F,\ G$ are four Thiele-Innes coefficients, $P$ and $t_p$\footnote{The reference epoch of Gaia DR3 is JD 2457389.} represent the orbital period and the time of periastron, $e$ and $\varpi$ are the orbital eccentricity and stellar parallax. We transform the correlation matrix of the Gaia two-body solution to a covariance matrix $\mathcal{C}$ using \code{nsstools}\footnote{\url{ https://www.cosmos.esa.int/web/gaia/dr3-nss-tools}} \citep{Halbwachs2023}. The Gaia likelihood function $\mathcal{L}_{\rm Gaia,~two\,body}$ can thus be established as:

\begin{equation}\label{likelihood_gaia_two_body}
  \mathcal{L}_{\rm Gaia,~two\,body}= \frac{1}{\sqrt{(2\pi)^{8}|{\rm det}(\mathcal{C})|}}{\rm exp}\left[{-\frac{\Theta^{T}\mathcal{C}^{-1}\Theta}{2}}\right],
\end{equation}
where $\Theta$ is the eight-element vector describing the deviation between the two-body solution and a set of parameters given in Equation~\ref{two_body_param}. 

In terms of the RVs, we utilize the \code{radvel} package \citep{Fulton2018} to build the Keplerian model. The RV likelihood function is taken to be 
\begin{equation} \label{rv_likelihood}
\begin{split}
  \mathcal{L}_{\rm RV}
  &
  =\prod\limits_{j=1}^{N_{\rm RV}}\prod\limits_{k=1}^{N_{\rm inst}}
    \frac{1}{\sqrt{2\pi(\sigma_{j,k}^2+\sigma_{k}^2)}} \times \\ 
  & \quad
  \exp\!\left( -\frac{(v_{j,k}-\hat{v}_{j,k}-\gamma_k)^2}
    {2(\sigma_{j,k}^2+\sigma_{k}^2)} \right)~,
\end{split}
\end{equation}
where $N_{\rm RV}$ and $N_{\rm inst}=2$ are the number of RV measurements and instruments (i.e., HARPS-pre and HARPS-post), $\gamma_k$ denotes the RV offset for the $k$-th instrument, while $\sigma_{k}$ represents the corresponding jitter to account for unmodeled stellar or instrumental noise, $v_{j,k}$ and $\hat{v}_{j,k}$ are the observed and predicted RVs while $\sigma_{j,k}$ is the observed RV uncertainty. We adopt the convention that the inclination $i$ ranges from 0 \(^\circ\) to 180\(^\circ\), following the convention used by Gaia and \cite{Winn2022}, with the corresponding ascending-node convention used consistently in the RV model.

We next carry out the fit using the joint likelihood function $\mathcal{L}_{\rm joint}=\mathcal{L}_{\rm Gaia,\ two\ body}\times \mathcal{L}_{\rm RV}$. The model has a total of 13 free parameters, including stellar mass $M_\ast$, companion mass $M_c$, two parameters relevant to the orbital eccentricity and argument of periastron ($e \cos \omega$ and $e \sin \omega$), the cosine of orbital inclination $\cos i$, longitude of ascending node $\Omega$, orbital period $P$, time of periastron $t_p$, stellar parallax $\varpi$, as well as two RV offsets $\gamma_{\rm pre}$, $\gamma_{\rm post}$ and two RV jitters $\sigma_{\rm pre}$, $\sigma_{\rm post}$. The parameters above can be directly converted to four ``Campbell'' orbital elements: the semi-major axis in angular units ($a_{0}$), the argument of periastron ($\omega$), the orbital inclination ($i$) as well as the longitude of ascending node ($\Omega$), in order to compute the Thiele-Innes coefficients, and the eight-element deviation vector $\Theta$ in Equation~\ref{likelihood_gaia_two_body}. 

Since the companion mass of \tar B is well below the hydrogen burning limit based on the mass reported by \cite{Feng2022}, the flux ratio $\varepsilon$ between the companion and the host is less than $2.2\times10^{-5}$ according to \cite{mamajek2013ApJS}. This value is negligible and we therefore fix the flux ratio $\varepsilon$ at 0. We adopt wide non-informative uniform priors on all parameters, except for the stellar mass $M_\ast$, where we place a Gaussian prior based on our findings from the spectroscopic analysis (Section~\ref{stellar_properties}). We employ the \code{emcee} package \citep{Foreman2013} to sample the posteriors of all parameters. We initialize 100 walkers, and all of them run for 500000 steps. We use two times of the maximum auto correlation time as burnt-in samples. 
We summarize the priors we adopted, and the best-fit results in Table~\ref{allpriors}. Figure~\ref{fig:rv_fit} illustrates the RV data of \tar\ along with the Keplerian model from the joint analysis. The posterior distributions of the fitted parameters are presented in Figure~\ref{fig:two_body_solution}.

\subsection{Joint Fit of RVs and HIP+GDR2+GDR3 Astrometry}\label{joint_fit_2}

For comparison, we also perform a joint modeling of the RVs and Hipparcos-Gaia astrometry to constrain the 3-D orbit and determine the true mass of \tar B, following the method developed by \cite{Feng2023} and modified by \cite{Xiao2024}. 
The primary fitted parameters include seven orbital elements (orbital period $P$, RV semi-amplitude $K$, two parameters relevant to eccentricity and argument of periastron $\sqrt{e}\cos \omega$ and $\sqrt{e}\sin w$, the cosine of orbital inclination $\cos i$, longitude of ascending node $\Omega$, mean anomaly $M_{0}$ at epoch 2016.0 (JD2,457,389.0)), five astrometric offsets ($\Delta \alpha*,\ \Delta \delta,\ \Delta \varpi,\ \Delta \mu_{\alpha*},\ \Delta \mu_{\delta}$) of barycenter relative to GDR3, two RV offsets $\gamma_{\rm pre}$, $\gamma_{\rm post}$ and the corresponding jitter terms $\sigma_{\rm pre}$, $\sigma_{\rm post}$, as well as an astrometric jitter $J_{\rm hip}$ for Hipparcos IAD
, and an error inflation factor $S_{\rm gaia}$ for Gaia astrometry. 


The likelihood function for Hipparcos IAD can be calculated as
\begin{equation}
  \mathcal{L}_{\rm hip}=\prod\limits_{j=1}^{N_{\rm IAD}}\frac{1}{\sqrt{2\pi(\sigma_{j}^2+J_{\rm hip}^2)}}
  \exp\!\left( -\frac{(\xi_j-\hat{\xi}_j)^2}{2(\sigma_{j}^2+J_{\rm hip}^2)} \right)~,
\end{equation}
where $N_{\rm IAD}$ is the number of epochs of Hipparcos IAD, $\sigma_j$ is the individual measurement uncertainty, and $J_{\rm hip}$ is the jitter term. 
Observed and predicted 1-D abscissae data are denoted with $\xi_j$ and $\hat{\xi}_j$, respectively. 
The abscissae $\hat{\xi}_j$ is synthesized by projecting the combined motion of the system's barycenter, the stellar reflex motion, and the parallax motion onto the 1-D scan direction of Hipparcos. 
Moreover, the offset between Hippacos astrometry and the astrometry propagated from the GDR3 epoch to the Hipparcos epoch are also incorporated into the modeling of $\hat{\xi}_j$.

In addition, we use multiple Gaia data releases by simulating the Gaia epoch data with Gaia Observing Schedule Tool\footnote{\url{https://gaia.esac.esa.int/gost/}} (GOST), since the Gaia IAD is not available until December in 2026.
The likelihood function for Gaia Data Release 2 (GDR2) and Gaia Data Release 3 (GDR3) can be written as
\begin{equation}
\begin{split}
  \mathcal{L}_{\rm gaia}=
  &
  \prod\limits_{k=1}^{N_{\rm DR}}\frac{1}{\sqrt{(2\pi)^5|\Sigma_k(S_{\rm gaia}^2)|}}
  \times\\
  &
  \exp\!\left( -\frac{1}{2}(\Delta\vec{\iota}_k-\Delta\hat{\vec{\iota}}_k)^{T}[\Sigma_k(S_{\rm gaia}^2)]^{-1}(\Delta\vec{\iota}_k-\Delta\hat{\vec{\iota}}_k)    \right)~,
\end{split}
\end{equation}
where $N_{\rm DR}=2$ represents the number of Gaia data releases (i.e., GDR2 and GDR3), $\Sigma_k$ is the covariance provided by Gaia catalog, and $S_{\rm gaia}$ is the error inflation factor. $\Delta\vec{\iota}_k$ is the five-parameter astrometric vector for catalog $k$ expressed relative to GDR3, e.g.,
\begin{equation}
\begin{split}
\Delta\vec{\iota}_k\equiv
&
\,((\alpha_k-\alpha_{\rm DR3})\,{\rm cos\delta_k},\,\delta_k-\delta_{\rm DR3},\,\\
&
\varpi_k-\varpi_{\rm DR3},\,\mu_{\alpha k}-\mu_{\alpha \rm DR3},\,\mu_{\delta k}-\mu_{\delta \rm DR3}),
\end{split}
\end{equation}
which can be obtained by fitting the standard five-parameter astrometric model to the simulated abscissae.
We implement the frame rotation calibration between GDR2 and GDR3 epoch data recommended by \cite{Feng2024}\footnote{\url{https://github.com/ruiyicheng/Download_HIP_Gaia_GOST}}.

\begin{figure*}[htb]
\centering
\includegraphics[width=1.0\textwidth]{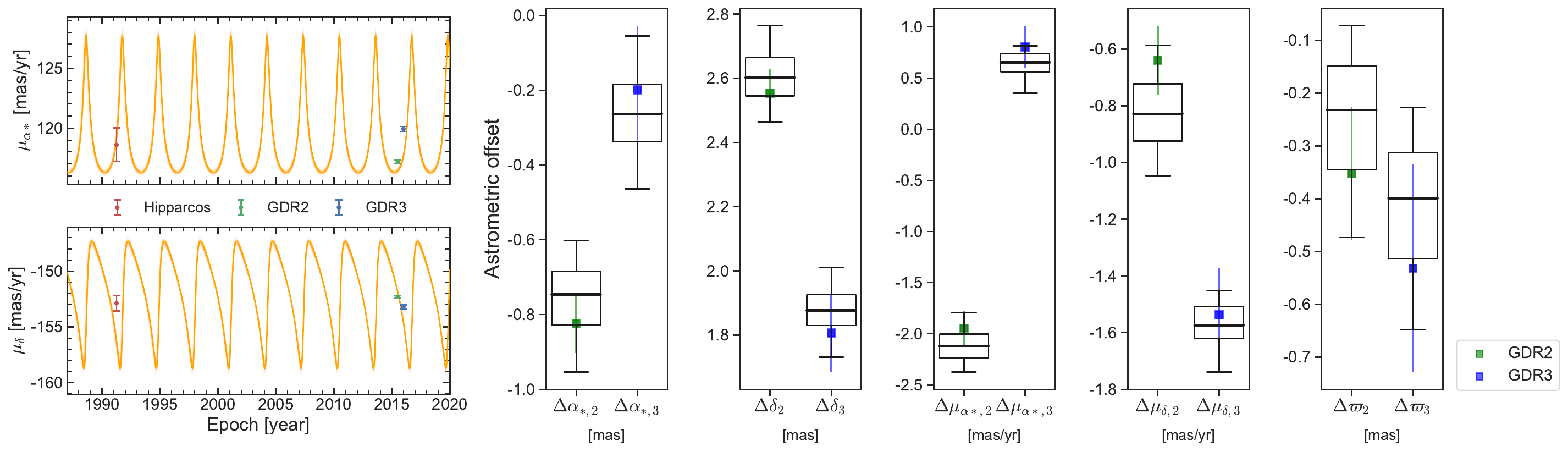}
\caption{Left panel: Proper-motion evolution of HD 203771 predicted by the joint RV+Hip+GDR2+GDR3 fit. 
The thick orange curves show the instantaneous model proper motion, including the fitted barycentric proper motion and time-dependent reflex proper motion induced by HD 203771 B; thin curves show orbital solutions randomly selected from the MCMC posterior distributions. 
The red, green, and blue points denote the Hipparcos, Gaia DR2, and Gaia DR3 catalog proper motions with their uncertainties. 
Right panel: Comparison between the model-predicted and catalog values of the five astrometric offsets for GDR2 and GDR3. 
The barycentric motion of the HD 203771 system has been subtracted from both the Gaia catalog data (colored squares) and the model predictions (box plots). 
The catalog error bars have been scaled by the fitted Gaia error inflation factor. 
The central thick line, edges of the box and whiskers indicate the median, $1\sigma$ and $3\sigma$ posterior credible intervals, respectively. 
Subscripts 2 and 3 denote GDR2 and GDR3.}
\label{fig:pmfit+boxplot_hip_gaia_joint}
\end{figure*}

Similar as above, we place uniform priors for all fitting parameters, except for stellar mass, where we assign the same Gaussian prior as in Section~\ref{two_body_RV_fit}. 
With the total likelihood $\mathcal{L} = \mathcal{L}_{\rm hip} \times \mathcal{L}_{\rm gaia}\times \mathcal{L}_{\rm RV} $, where the RV likelihood function is the same as Equation~\ref{rv_likelihood}, we derive the orbital solution by sampling the posterior via the parallel-tempering MCMC sampler \code{ptemcee} \citep{Vousden2016}.  
We employ 25 temperatures, 100 walkers, and 50,000 steps per chain to generate posterior distributions for all the fitting parameters, with the first 20000 steps being discarded as burn-in. We summarize the best-fit orbital parameters in Table~\ref{allpriors}. Figure \ref{fig:pmfit+boxplot_hip_gaia_joint} depicts the fitting of proper motion and five astrometric offsets for HD 203771. We show the posterior distributions of the fitting parameters and derived parameters for HD 203771 in Figure \ref{fig:cornerplots_RV+Hip+GDR2+GDR3}. In the next section we compare the outputs of the two methodologies, and the result from the previous work \cite{Feng2022}.


\begin{figure}[htb]
\centering
\includegraphics[width=0.49\textwidth]{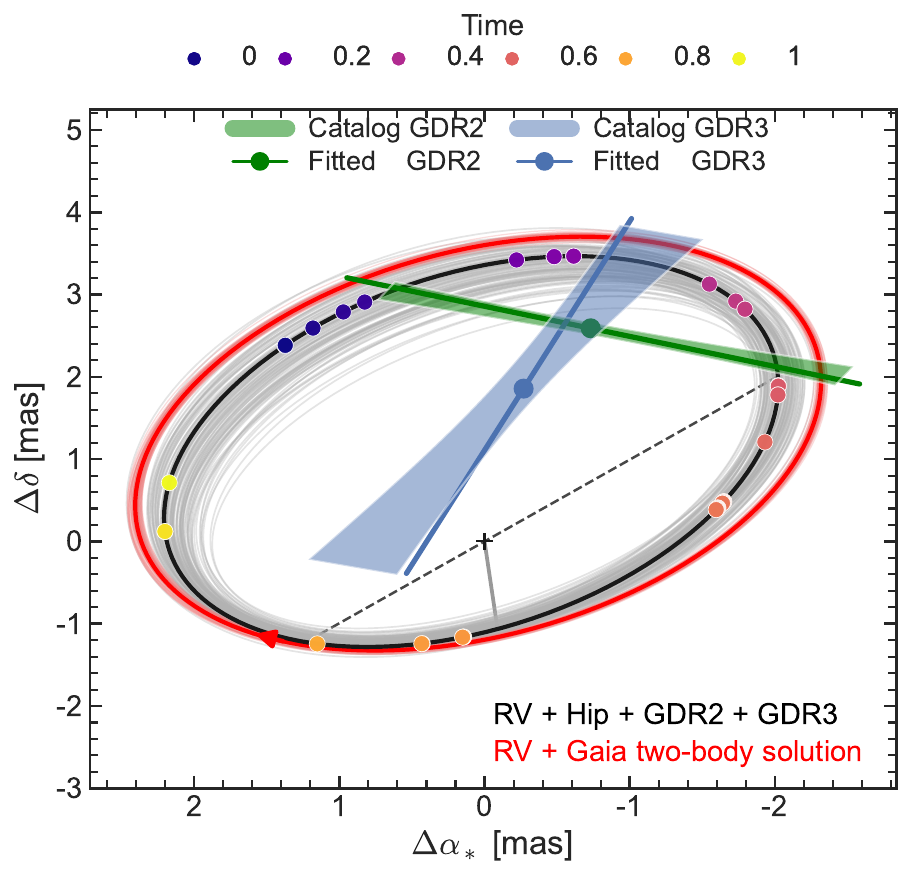}
\caption{Comparison of the HD 203771 astrometric orbit from the RV+Gaia DR3 two-body solution (red) and the joint RV+Hip+GDR2+GDR3 fit (black). 
The + sign marks the position of barycenter. 
(a) RV + Gaia DR3 two-body solution: The thick red curve represents the best-fit astrometric orbit, while the thin red curves represent 100 randomly selected posterior orbits. The arrow marks the direction of orbital motion. 
(b) RV+Hip+GDR2+GDR3: The thick black curve shows the best-fit orbit, while thin gray curves show 100 posterior draws. 
The green and blue shaded regions indicate the barycentric-motion-subtracted uncertainty envelopes for GDR2 and GDR3.
The green and blue circles mark companion-induced position offsets at the GDR2 and GDR3 reference epochs, and the associated solid lines show their evolution due to the fitted proper-motion offsets. 
The colored points along the black orbit show the model-predicted astrometric positions at the Gaia scanning epochs and are color-coded by normalized observing time, increasing from purple to yellow. }
\label{fig:orbits}
\end{figure}

\begin{table*}\scriptsize
    \centering
    {\renewcommand{\arraystretch}{1.05}
    \caption{Parameter priors and best-fits in two joint models. $\mathcal{U}$(a, b) stands for a uniform prior between $a$ and $b$. $\mathcal{LU}$(a\ , \ b) is a log-uniform prior between $a$ and $b$. $\mathcal{N}$($\mu\ ,\ \sigma^{2}$) is a normal prior with mean $\mu$ and standard deviation $\sigma$, while $\mathcal{TN}$ ($\mu$, $\sigma^{2}$, $a$\ ,\ $b$) means a truncated normal prior ranging from $a$ to $b$.} 
    \begin{tabular}{lcccr}
        \hline\hline
        Parameter       &Prior &Best-fit &Best-fit    &Description\\
        & &RV+Gaia  &RV+HIP+GDR2+GDR3 &\\
        & &Two-body Orbital Solution &Absolute Astrometry \\\hline
        \it{Fitted Parameters}\\
        $M_\ast$ ($M_\odot$) &$\mathcal{N}$ ($0.79$\ ,\ $0.08^{2}$)  &$0.79\pm0.06$ &$0.79\pm0.08$ &Stellar mass\\
        $P$ (days)  &$\mathcal{U}$ ($1000$\ ,\ $1300$)  &$1140.366^{+0.915}_{-0.879}$ &$1140.022^{+1.456}_{-1.139}$
        &Orbital period\\
        $t_{p}$ (JD-2457389) &$\mathcal{U}$ ($0$\ ,\ $500$) &$256.085^{+1.246}_{-1.534}$ &$256.774^{+1.482}_{-2.111}$ $^{\dag}$ &Time of periastron\\
        $M_{c}$ ($M_{\mathrm{J}}$) &$\mathcal{U}$ ($0$\ ,\ $150$) &$74.20^{+3.46}_{-3.64}$ &$69.03^{+4.99}_{-4.95}$ $^{\dag}$ &Companion mass\\
        $e\cos \omega$ &$\mathcal{U}$ ($-1$\ ,\ $1$) &$0.236^{+0.005}_{-0.006}$ &$\cdots$ &Parametrization for $e$ and $\omega$\\
        $e\sin \omega$ &$\mathcal{U}$ ($-1$\ ,\ $1$) &$-0.443^{+0.004}_{-0.005}$ &$\cdots$ &Parametrization for $e$ and $\omega$\\
        
        $\sqrt{e} \cos \omega$ &$\mathcal{U}$ ($-1$\ ,\ $1$) &$\cdots$ &$0.338^{+0.008}_{-0.009}$ &Parametrization for $e$ and $\omega$\\
        $\sqrt{e} \sin \omega$ &$\mathcal{U}$ ($-1$\ ,\ $1$) &$\cdots$ &$-0.627^{+0.005}_{-0.005}$ &Parametrization for $e$ and $\omega$\\
        
        $\cos i$ &$\mathcal{U}$ ($-1$\ ,\ $1$)  &$-0.689\pm0.003$ &$-0.637^{+0.028}_{-0.024}$ 
        &Cosine of the orbital inclination\\
        $\varpi$ (mas) &$\mathcal{U}$ ($17$\ ,\ $19$) &$18.488\pm0.019$ &$\cdots$ &Stellar parallax\\
        $\Omega$ (rad) &$\mathcal{U}$ ($-\pi$\ ,\ $\pi$) &$2.266\pm0.009$ &$2.320 \pm 0.036$ &Longitude of ascending node \\
        $M_0$ (rad) &$\mathcal{U}$ ($0$\ ,\ $2\pi$) &$4.873^{+0.008}_{-0.008}$ $^{\dag}$ &$4.868^{+0.012}_{-0.008}$ &Mean anomaly at 2457389\\
        $\Delta \alpha \ast$ (mas) &$\mathcal{U}$ ($-10^{6}$\ ,\ $10^{6}$) &$\cdots$ &$-0.242^{+0.102}_{-0.099}$ &Offset of RA in GDR3\\
        $\Delta \delta$ (mas) &$\mathcal{U}$ ($-10^{6}$\ ,\ $10^{6}$) &$\cdots$ &$1.710^{+0.109}_{-0.109}$ &Offset of Dec in GDR3\\
        $\Delta \mu_{\alpha\ast}$ (mas~yr$^{-1}$) &$\mathcal{U}$ ($-10^{6}$\ ,\ $10^{6}$) &$\cdots$ &$0.837^{+0.050}_{-0.050}$ &Offset of proper motion in RA in GDR3\\
        $\Delta \mu_{\delta}$ (mas~yr$^{-1}$) &$\mathcal{U}$ ($-10^{6}$\ ,\ $10^{6}$) &$\cdots$ &$-1.536^{+0.030}_{-0.030}$ &Offset of proper motion in Dec in GDR3\\
        $\Delta \varpi$ (mas) &$\mathcal{U}$ ($-10^{6}$\ ,\ $10^{6}$) &$\cdots$ &$-0.490^{+0.103}_{-0.103}$ &Offset of parallax in GDR3\\
        $K$ (m~s$^{-1}$) &$\mathcal{LU}$ ($-10$\ ,\ $10$) &$\cdots$  &$1335.437^{+5.493}_{-5.667}$ &RV semi-amplitude\\
        $\gamma_{\rm pre}$ (m~s$^{-1}$)  &$\mathcal{U}$ ($-500$\ ,\ $500$) &$-275.3^{+9.1}_{-8.0}$ &$-282.8^{+12.4}_{-10.7}$ &RV offset of HARPS-pre\\
        $\gamma_{\rm post}$ (m~s$^{-1}$) &$\mathcal{U}$ ($-500$\ ,\ $500$) &$404.8^{+4.2}_{-4.1}$ &$404.5^{+4.3}_{-4.5}$ &RV offset of HARPS-post\\
        $\sigma_{\rm pre}$ (m~s$^{-1}$) &$\mathcal{U}$ ($0$\ ,\ $30$) &$2.1^{+3.5}_{-1.5}$ &$3.0^{+12.8}_{-2.3}$ &RV jitter of HARPS-pre\\
        $\sigma_{\rm post}$ (m~s$^{-1}$) &$\mathcal{U}$ ($0$\ ,\ $30$) &$13.3^{+5.0}_{-3.2}$ &$12.3^{+4.3}_{-2.7}$ &RV jitter of HARPS-post\\
        $J_{\rm hip}$ (mas) &$\mathcal{U}$ ($10^{-6}$\ ,\ $10^6$) &$\cdots$ &$0.968^{+0.976}_{-0.674}$ &Jitter for Hipparcos\\
        $S_{\rm gaia}$ &$\mathcal{TN}$ ($1.0$, \ $0.1^2$; $1, \ 2$) &$\cdots$ &$1.072^{+0.074}_{-0.050}$ &Gaia error inflation factor \\\hline
        \it{Derived parameters}\\
        $a$ (AU) &$\cdots$ &$2.035^{+0.045}_{-0.047}$ &$2.026^{+0.065}_{-0.068}$ &Semi-major axis\\
        $i$ (deg) &$\cdots$ &$133.55^{+0.24}_{-0.23}$ &$129.55^{+1.83}_{-2.06}$ &Orbital inclination\\
        $q$ &$\cdots$ &$0.089 ^{+0.003}_{-0.002}$ &$0.084^{+0.004}_{-0.004}$ &Companion-to-star mass ratio\\
        $e$ &$\cdots$ &$0.502^{+0.005}_{-0.005}$ &$0.507^{+0.006}_{-0.006}$ &Orbital eccentricity\\
        $\omega$ (rad) &$\cdots$ &$-1.081^{+0.011}_{-0.010}$ &$-1.076^{+0.011}_{-0.013}$ &Argument of periapsis\\\hline
    \label{allpriors}    
    \end{tabular}}
    \begin{tablenotes}
    \item[1]  [1]\ Apart from five derived parameters shown below, results marked by $\dag$ on the top right are also the derived values, which are not directly fitted in the corresponding model but calculated from fitted parameters assuming mutual independence.
    \end{tablenotes}
\end{table*}


\subsection{HD 203771B as an Intermediate-orbit Brown Dwarf}



\cite{Feng2022} first measured the mass for \tar B ($M_c=56.29^{+9.52}_{-5.21}\ M_{\mathrm{J}}$) through a joint analysis of Gaia EDR3 and Hipparcos catalog astrometry together with four RVs from HARPS-pre, identifying the companion as a brown dwarf. We refine the companion mass measurements through the two independent methods, leading to $74.20_{-3.64}^{+3.46} M_{\mathrm{J}}$ obtained through RV + Gaia two-body solution analysis as well as $69.03^{+4.99}_{-4.95} M_{\mathrm{J}}$ from RV + Hip + GDR2 + GDR3 catalog astrometry joint fit. In addition to including more RVs in this study, the primary difference between orbital solutions in two works lies in the orbital inclination. \cite{Feng2022} obtained an inclination of $106.57_{-11.10}^{+9.26}$\(^\circ\). 
In contrast, a combination of the two-body solution with RVs refines the inclination angle to $133.55^{+0.24}_{-0.23}$\(^\circ\), while the Hipparcos IAD and two-epoch Gaia astrometry plus RVs joint analysis results in an inclination of $129.55^{+1.83}_{-2.06}$\(^\circ\). 
Figure \ref{fig:orbits} presents the astrometric orbits derived from two aforementioned methodologies, which are consistent with each other within $1\sigma$.
Nevertheless, the RV+Gaia two-body solution yields tighter constraints on the orbital parameters, leading to a more precise determination to the physical properties of the companion. 


Both methods we use benefit from incorporating time-series astrometric information. This advantage is particularly important for systems with orbital periods comparable to or shorter than the $\sim$1000-day observational baseline of the Hipparcos and Gaia catalogs, for which catalog-level astrometric solutions capture only part of the orbital motion and therefore provide weaker orbital constraints \citep{Feng2022}. 
At present, mission-wide time-series astrometric data are publicly available only through the Hipparcos IAD, although for a limited sample of targets \citep{vanLeeuwen2007}. While Gaia has not yet released its individual time-series astrometric measurements, the Gaia DR3 Non-Single Star (NSS) catalog provides two-body solutions derived from these observations and has already played an important role in determining the true masses of brown dwarf companions \citep[e.g.,][]{Unger2023,Barbato2026}. 

\begin{figure}[htb]
    \centering
    \includegraphics[width=0.49\textwidth]{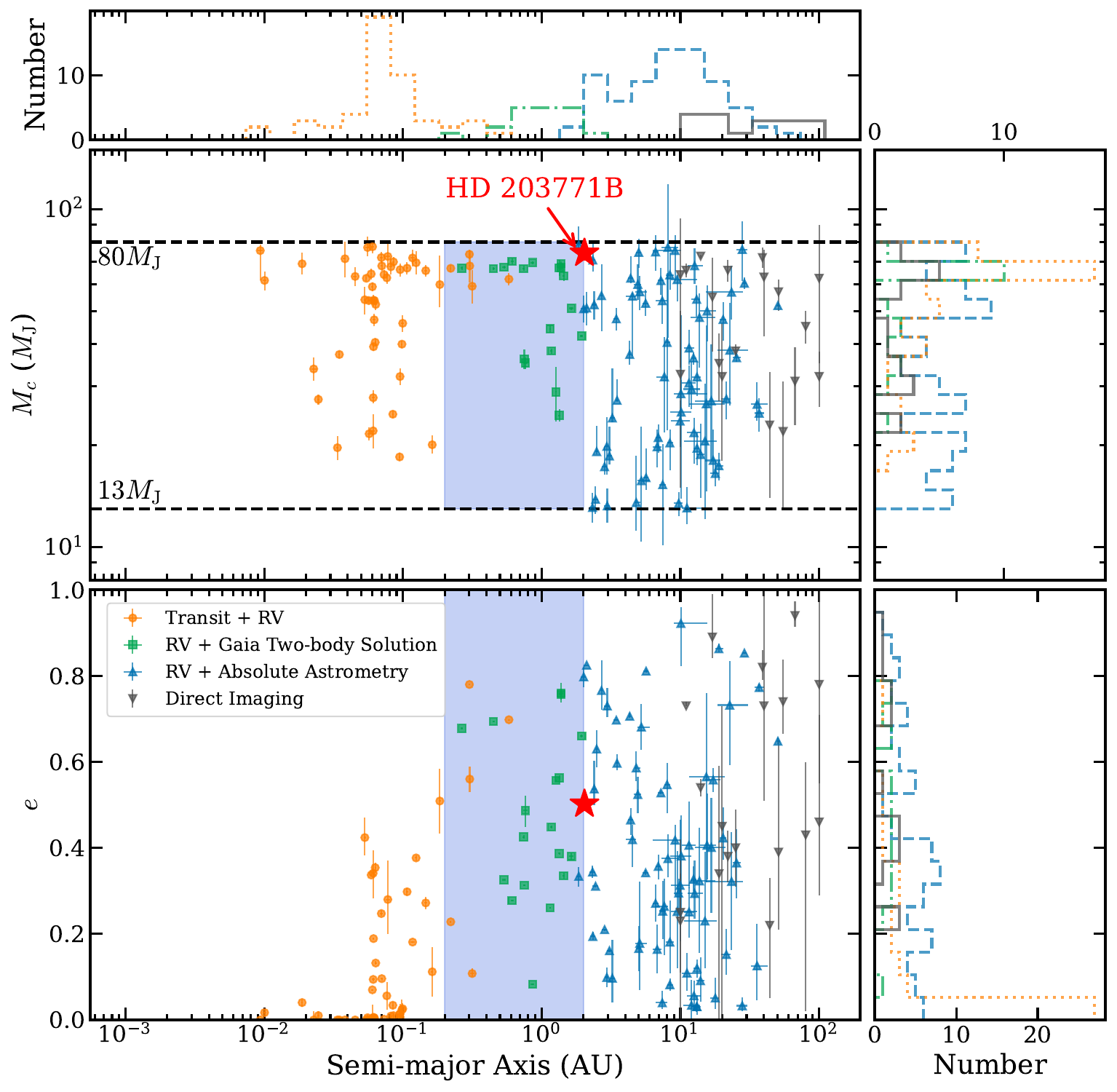}
    \caption{The mass versus semi-major axis and eccentricity versus semi-major axis distributions of 159 bounded brown dwarf with absolute mass measurements, colored by different detection techniques. The full sample includes transiting brown dwarfs \citep{Gan2025}, brown dwarfs characterized through RV+Gaia two-body solutions \citep{Barbato2026,Unger2023}, and RV+Hipparcos+Gaia astrometry \citep{Feng2022}, as well as brown dwarfs detected by direct imaging \citep{Bowler2020}. The horizontal black dashed lines represent the empirical lower ($13\ M_{\mathrm{J}}$) and upper ($80\ M_{\mathrm{J}}$) mass limits of the brown dwarf regime, while the shaded area highlights the relatively sparse region that is now being filled up by astrometry. Our target \tar B is marked as a red star.}
    \label{fig:massperiod}
\end{figure}

\tar B joins the small but growing sample of intermediate-orbit brown dwarfs with absolute mass measurements locating at about 2 AU. As shown in Figure \ref{fig:massperiod}, the current brown dwarf sample exhibits a noticeable deficit of companions with semi-major axes between approximately 0.2 and 2 AU (the blue shaded region, corresponding to orbital periods of roughly 30–1000 days). The upper boundary of the gap coincides with the regime in which Gaia–Hipparcos catalog astrometry becomes less effective for orbit determination as discussed above. 
The lower boundary of this gap is likely influenced by the declining sensitivity of transit surveys to longer-period companions. As orbital period increases, the geometric transit probability decreases and it becomes more difficult to obtain multiple transits within the finite survey lifetime \citep{Sullivan2015}. Consequently, the transit method is strongly biased toward detecting short-period systems, resulting in a rapid decline in the number of transiting companions at longer orbital periods \citep{Jiang2019}. The forthcoming Gaia DR4 time-series astrometric data, which span 5.5-year (or 66 months) baseline, will bridge this observational gap and remedy the situation. 

\section{The Eccentricity Distribution of Brown Dwarfs}\label{eccentricity}

Combining the bounded brown dwarfs detected by transit+RV, RV+astrometry and imaging methods that have absolute mass measurements (see Figure~\ref{fig:massperiod}), we construct a sample of 159 brown dwarfs with semi-major axis spanning the range from 0.01 to 100 AU, enabling us to investigate the eccentricity distributions at different orbital separations. 

We first group the full brown dwarf sample into three subsamples with semi-major axes of $0.1<a<1$ AU, $1<a<20$ AU, and $a>20$ AU, respectively. The inner threshold 0.1 AU, which roughly corresponds to an orbit period cut at about 10 days, is selected to minimize the influence of stellar tidal effect that will erase the orbital eccentricity of brown dwarfs \citep[e.g.,][]{Gan2025}. Therefore, we adopt 126 brown dwarfs with $a>0.1$ AU in the following statistical analysis, while the remaining 33 systems are excluded. The other two limits are chosen somewhat arbitrarily, but roughly marking the boundaries between transit and astrometry discovery space, as well as astrometry and imaging detections as shown in Figure~\ref{fig:massperiod}. 

\begin{figure*}[htb]
    \centering
    \includegraphics[width=0.99\textwidth]{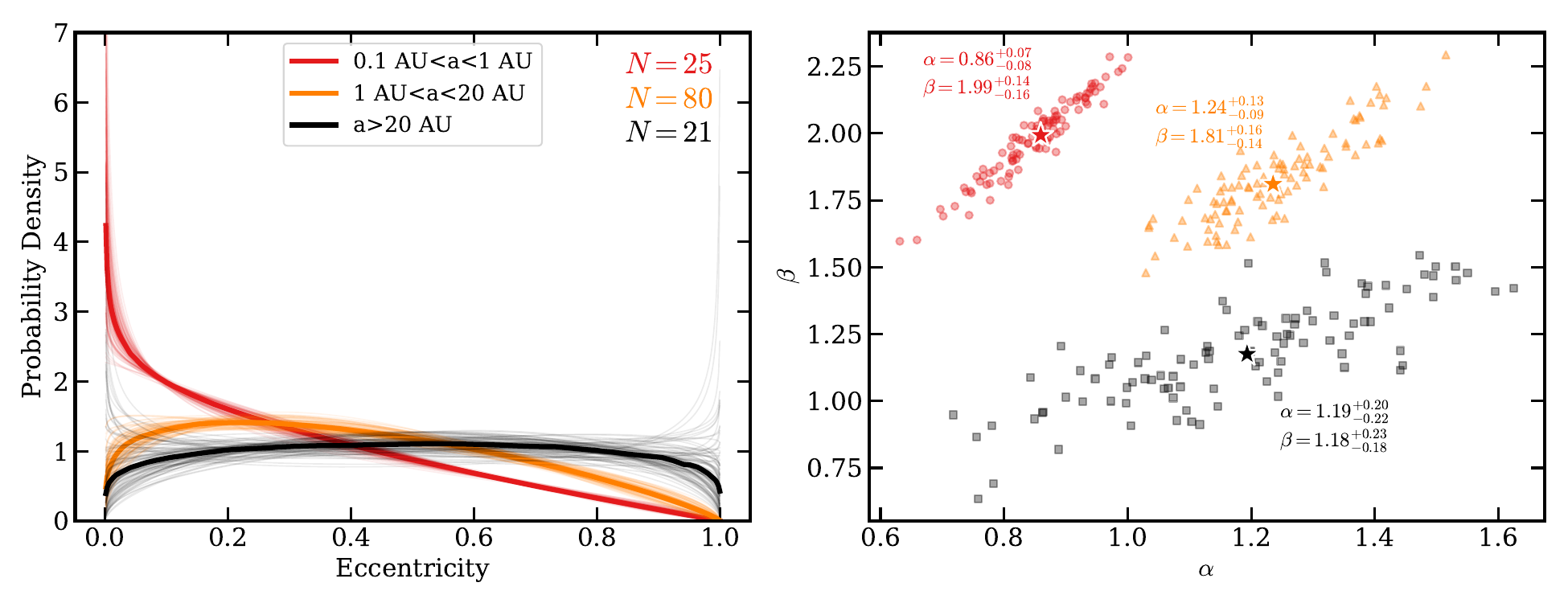}
    \caption{Left panel: Best-fit beta distributions for 100 simulated eccentricity data sets of each brown dwarf subsample, divided according to semi-major axis. Each simulated data set is generated by randomly sampling the measured eccentricities from truncated Gaussian distributions defined by the reported measurement uncertainties. The thin curves represent the best-fit beta distributions for the individual simulated data sets, while the bold curves show the point-wise mean of the 100 fitted probability density curves for each subsample. The size of each subsample is shown on the upper right corner. Right panel: Scatter plot of the corresponding best-fit beta distribution shape parameter pairs $(\alpha,\beta)$ for the 100 simulated data sets of each subsample. The annotated values indicate the mean and standard deviation of the fitted shape parameters. 
    }
    \label{fig:eccdist}
\end{figure*}

For each subsample, we opt to fit the eccentricities using a Beta distribution with the help of \code{scipy.stats.beta} \citep{Virtanen2020} through the least $\chi^{2}$ method. We adopt a Beta distribution because it is naturally defined on the interval $(0,1)$ and can represent a variety of distribution shapes, including monotonically decreasing and increasing distributions. While Rayleigh distribution, which is another model widely adopted in previous works \citep[e.g.,][]{Wu2025,Sagear2026}, cannot capture monotonic distributions. 
The Beta distribution is formulated as 

\begin{equation}
    f(e;\alpha, \beta)= \frac{\Gamma(\alpha+\beta)e^{\alpha-1}(1-e)^{\beta-1}}{\Gamma(\alpha)\Gamma(\beta)},\label{beta_euation}
\end{equation}
where $e$ represents the orbital eccentricity, $\alpha$ and $\beta$ are two parameters that determine the shape of the Beta distribution \citep{Kipping2013}, while $\Gamma$ represents the gamma function. In order to take eccentricity uncertainties into consideration, we utilize the Monte Carlo method to generate simulated data. For every system, we assume a truncated Gaussian distribution $\mathcal{TN}(e,\sigma_{e}^{2},0,1)$, where $e$ and $\sigma_{e}$ are the measured eccentricity and its uncertainty, and we randomly draw 100 sets of data. To be conservative, $\sigma_{e}$ is chosen as the larger one of the asymmetric eccentricity uncertainties. In this way, we fit 100 independent Beta distributions for each of the three brown dwarf populations with different semi-major ranges. Figure~\ref{fig:eccdist} presents the resulting probability density functions of the eccentricities of three brown dwarf populations from our 100 fits, along with the distributions of their corresponding shape parameters, $\alpha$ and $\beta$, shown in the right panel. 

Although we emphasize that our sample is limited and possible detection bias from different methods is not accounted for, we find that the three subsamples exhibit systematically different eccentricity distributions with shape parameters of [$\alpha=0.86^{+0.07}_{-0.08}$, $\beta=1.99^{+0.14}_{-0.16}$], [$\alpha=1.24^{+0.13}_{-0.09}$, $\beta=1.81^{+0.16}_{-0.14}$], and [$\alpha=1.19^{+0.20}_{-0.22}$, $\beta=1.18^{+0.23}_{-0.18}$] for inner-orbit, intermediate-orbit, and wide-orbit brown dwarfs. More importantly, we find that the peak of eccentricity distribution tends to gradually shift toward lower values with decreasing orbital semi-major axis. Note that the selection effects of different techniques are unlikely to significantly change our conclusions. The reason is that transit surveys are more likely to find eccentric-orbit objects as the transit probability scales with $1/\sqrt{1-e^{2}}$ \citep{Barnes2007transitprobability}, while the detection sensitivity on highly-eccentric-orbit ($e>0.6$) planets from RV surveys is lower compared with circular-orbit planets \citep{Cumming2004}. Such an expected result, however, conflicts with our findings from observations here: inner brown dwarfs have more circular orbits than those wide-orbit ones found by RV, astrometry and imaging. To demonstrate that the results are insensitive to the choices of semi-major axis cuts, we apply alternative wide-orbit cutoffs at 10 and 30 AU, leading to the results shown in Figure~\ref{fig:e10AU}. We find that the overall trend of decreasing eccentricity peaking toward smaller separations is preserved. We discuss one possible hypothesis below to explain such a phenomenon. 

The large eccentricities of brown dwarfs at exterior orbits beyond 20 AU are supposed to be due to the gravitational instability formation history in massive disks that inherit high eccentricities \citep{Bowler2020}. Multiple self-gravitating fragments form within the disk, and subsequent gravitational interactions among these fragments excite the eccentricities of forming brown dwarfs \citep{Hall2017,Stamatellos2009}. After formation, both semi-major axis and orbital eccentricity of the brown dwarf may be damped through disk friction during inward migration, because the disk is much more massive than the brown dwarf \citep{Ida2020}. Under this framework, one would naturally expect to see lower eccentricities when moving to the inner region of the disk. This scenario also predicts that the eccentricity damping and semi-major axis damping timescales are comparable when the eccentricity is high ($e \sim 0.5$--1). As the eccentricity decreases, however, eccentricity damping becomes more efficient than semi-major axis damping, consistent with the observational results shown in Figure~\ref{fig:eccdist}. Interestingly, stellar obliquity studies show that the majority of brown dwarf companions with determined obliquity exhibit well-aligned orbits within about $1\sigma$ \citep[e.g.,][]{Triaud2009,Triaud2013,Siverd2012,Zhou2019b,Ferreira2024,Giacalone2024,Brady2025,Doyle2025,Carmichael2026,Vowell2026}, indicating a coplanar quiescent migration history, which supports the disk migration route we propose here. 
While eccentricity damping during disk-driven migration has been widely discussed for planetary objects \citep{Papaloizou2000, Cresswell2006, Cresswell2008, Bitsch2010, Kley2012, Ida2020}, its effectiveness in the brown dwarf mass regime is not sufficiently explored. Nevertheless, our findings indicate that a similar mechanism may also contribute to the orbital evolution of brown dwarf companions, which require further investigations. Stellar obliquity studies on brown dwarfs companions, especially on long-period objects, and their distributions compared to giant planets and binary stars will help probe their evolutionary history \citep[e.g.,][]{Gan2026}. 

However, we note that there are a few caveats we do not account for. First, as brown dwarfs migrate into inner orbits, tidal effect may contribute to orbital circularization even before they enter the $a<0.1$ AU region, and the influences from tidal effect and disk friction are challenging to distinguish from current observations. Second, the truncated Gaussian distribution we adopted cannot capture non-Gaussian posterior shapes, which may induce additional uncertainties into the inferred eccentricity distribution. However, this effect is expected to be negligible in this work, as the lower and upper uncertainties of the eccentricities used in analysis are generally comparable. Finally, we note that the eccentricities compiled from the literature were derived using different observational datasets, orbital-fitting methods, and prior assumptions, which may introduce systematic heterogeneity into our population analysis. Given the diverse detection methods and observational limitations across the brown dwarf sample, constructing a fully homogeneous catalog spanning a broad range of semi-major axes is challenging. Ideally, the orbital parameters of all systems should be derived homogeneously using the same datasets, fitting methodology, and prior assumptions; however, such a uniform analysis is beyond the scope of this work.

\section{Conclusions}\label{conclusion}


In this work, we revisited \tar B, an intermediate orbit brown dwarf companion previously characterized by \cite{Feng2022}, using a joint fit of RV measurements and two different time-series astrometry based data sets respectively. Compared with the previous analysis with RV + Gaia-Hipparcos catalog astrometry, our method provides a more tightly constrained absolute mass for \tar B. Placing it into the broader population of known brown dwarf companions, we find that systems with intermediate orbital separations (approximately 0.2--2 AU) are relatively scarce among the current brown dwarf sample with absolute mass measurements.

We further investigated the eccentricity distributions of brown dwarf companions in three semi-major axis ranges (close: $0.1<a<1$\,AU, intermediate: $1<a<20$\,AU and $a>20$\,AU). We find that wide-orbit brown dwarfs preferentially exhibit higher orbital eccentricities, while the peak of the eccentricity distribution systematically shifts toward lower eccentricities with decreasing semi-major axis. This trend may reflect dynamical excitation through interactions among clumps formed by gravitational instability at wide separations, followed by eccentricity damping caused by dynamical friction during inward migration through a gaseous disk (a similar scenario was proposed by \citealt{Ida2020} for smaller mass ratios). The forthcoming Gaia DR4 time-series astrometry is expected to fill the current gap in the intermediate-separation brown dwarf population, enabling more detailed studies of the eccentricity distribution, and thus providing stronger observational constraints on the formation and orbital evolution of brown dwarfs in contrast to planetary companions.


\section{Acknowledgments}


We thank the anonymous referee for constructive comments that improve the quality of this work. L.R., T.G. and S.M. acknowledge support by the National Natural Science Foundation of China (grant No. 12133005). T.G. and E.P. acknowledge financial support from the Agencia Estatal de Investigaci\'on of the Ministerio de Ciencia e Innovaci\'on MCIN/AEI/10.13039/501100011033 and the ERDF “A way of making Europe” through projects PID2021-125627OB-C32 and PID2024-158486OB-C32. This work is supported by the European Union (ERC AdvG SPEAR, GA 101200674).

%

\vspace{5mm}
\facilities{HARPS, Hipparcos, Gaia}


\software{corner \citep{Foreman2016}, emcee \citep{Foreman2013}, ptemcee \citep{Vousden2016}, radvel \citep{Fulton2018}, scipy \citep{Virtanen2020}
          }




\appendix

\counterwithin{table}{section}
\counterwithin{figure}{section}

\section{HARPS radial velocities of \tar}

Table~\ref{RVtable} lists the HARPS RVs of \tar\ from the HARPS-RVBank \citep{Trifonov2020,HARPSRVBank2}.

\begin{table}[htb]
    \centering
    {\renewcommand{\arraystretch}{0.85}
    \caption{HARPS radial velocities for \tar\ from the HARPS-RVBank.}
    \begin{tabular}{cccc}
        \hline\hline
        JD       &RV (m~s$^{-1}$) &$\sigma_{\rm RV}$ (m~s$^{-1}$) &Inst. \\\hline
        2455013.917 	&-1076.19 	&1.26 	&HARPS-pre\\
2455090.649 	&-1211.37 	&0.84 	&HARPS-pre\\
2455465.684 	&1265.39 	&1.04 	&HARPS-pre\\
2456555.749 	&1350.07 	&1.22 	&HARPS-pre\\
2456556.593 	&1352.97 	&2.07 	&HARPS-pre\\\hline
2457936.716 	&945.66 	&1.03 	&HARPS-post\\
2458392.646 	&-314.26 	&1.23 	&HARPS-post\\
2458442.551 	&-409.76 	&1.40 	&HARPS-post\\
2458678.704 	&-446.60 	&1.82 	&HARPS-post\\
2458693.717 	&-325.12 	&1.86 	&HARPS-post\\
2458719.785 	&-48.88 	&1.89 	&HARPS-post\\
2459160.600 	&645.75 	&1.82 	&HARPS-post\\
2459507.622 	&-241.83 	&1.73 	&HARPS-post\\
2459508.559 	&-244.42 	&1.84 	&HARPS-post\\
2459521.594 	&-266.07 	&1.84 	&HARPS-post\\
2459534.531 	&-316.96 	&1.82 	&HARPS-post\\
2459548.537 	&-349.40 	&1.89 	&HARPS-post\\
         \hline
    \label{RVtable}
    \end{tabular}}
\end{table}

\section{The SED of \tar}

\begin{figure}[htb]
    \centering
    \includegraphics[width=0.49\textwidth]{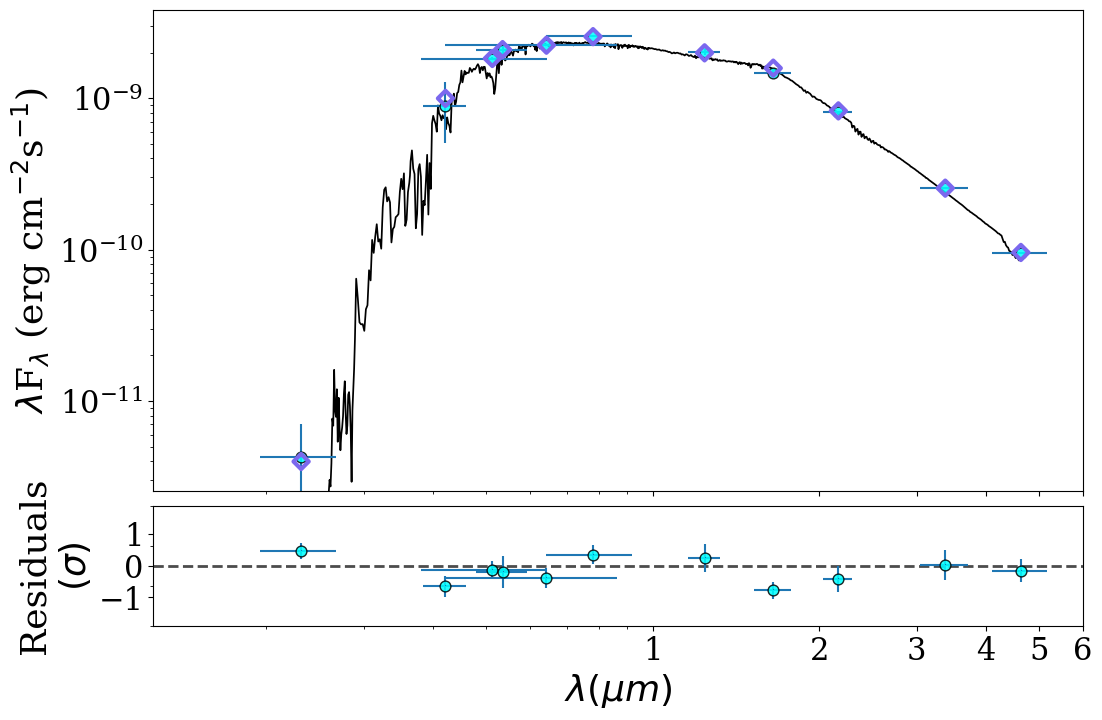}
    \caption{The SED of \tar\ together with the best-fit Kurucz model plotted in black. The residuals are shown in the bottom panel.}
    \label{fig:sed}
\end{figure}

Figure~\ref{fig:sed} shows the full stellar SED along with the best-fit Kurucz model.

\section{Posteriors of two joint-fit analyses}

\begin{figure*}[htb]
    \centering
    \includegraphics[width=0.98\linewidth]{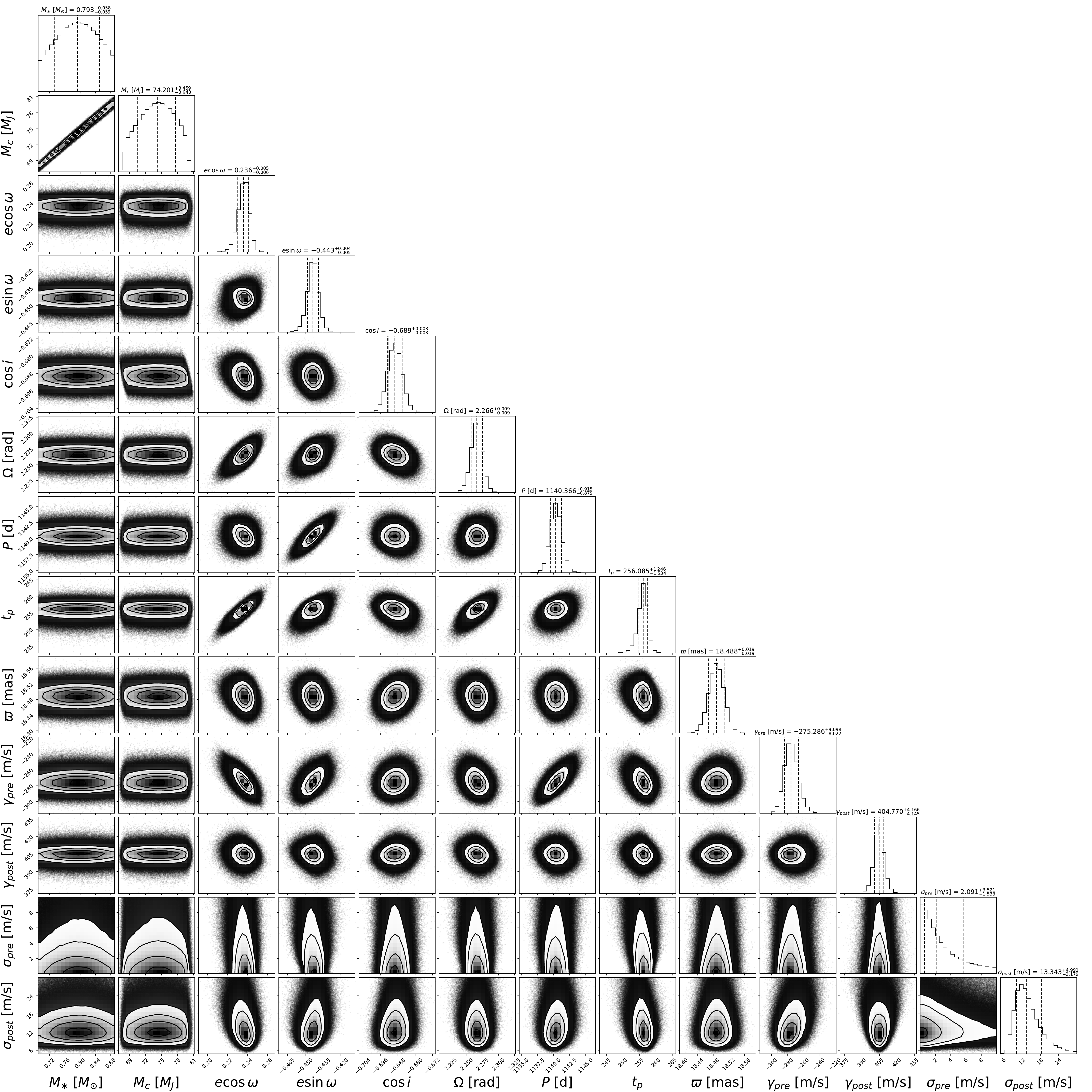}
    \caption{Posterior distributions of the fitting parameters from RV + Gaia two-body solution joint fit for HD 203771 system. } 
    \label{fig:two_body_solution}
\end{figure*}

\begin{figure*}[htb]
    \centering
    \includegraphics[width=0.98 \linewidth]{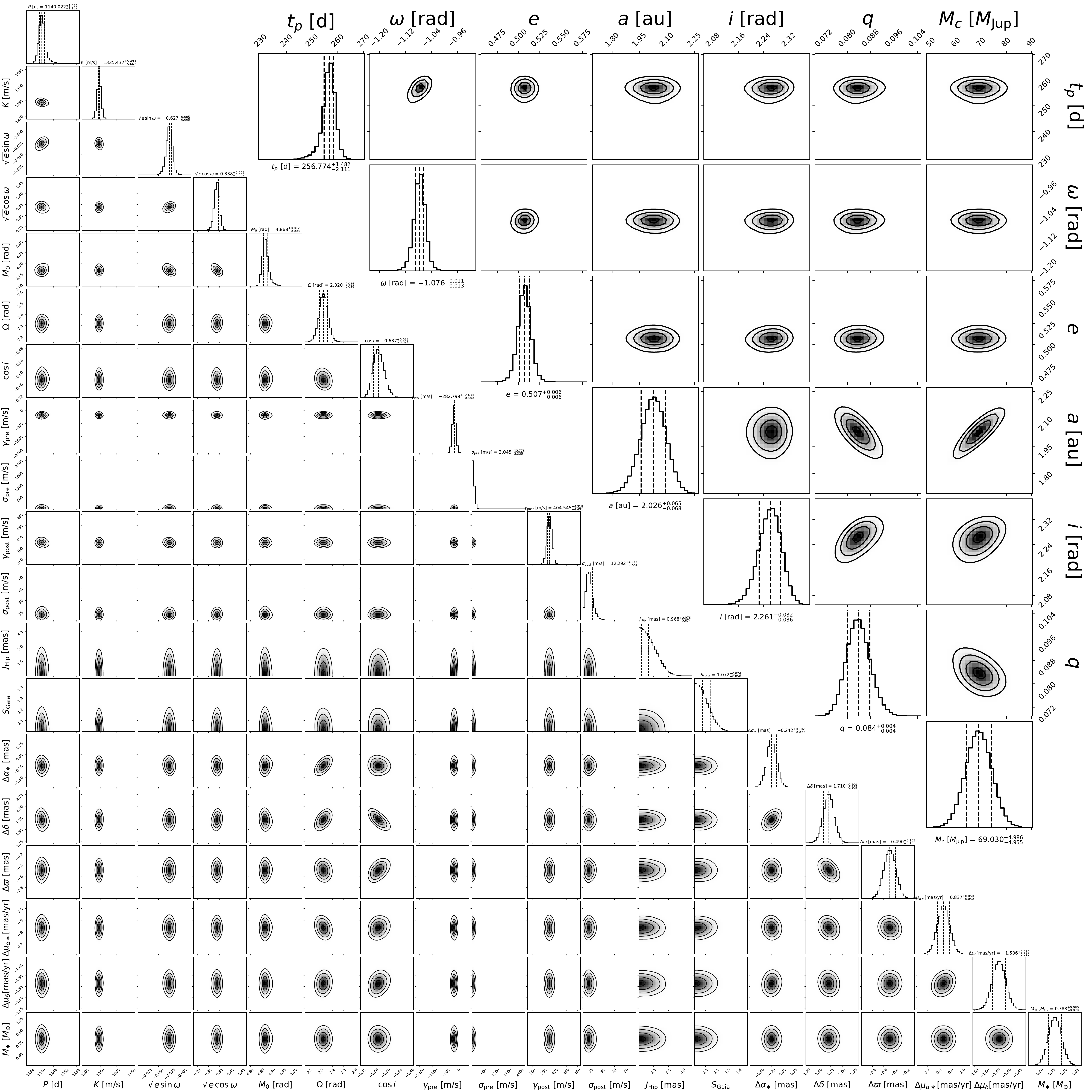}
    \caption{Posterior distributions of the fitting parameters (lower left) and derived parameters (upper right) from RV + Hip + GDR2 + GDR3 joint fit for HD 203771 system.}
    \label{fig:cornerplots_RV+Hip+GDR2+GDR3}
\end{figure*}

Figures~\ref{fig:two_body_solution} and \ref{fig:cornerplots_RV+Hip+GDR2+GDR3} present the posterior distributions of two different joint models: RV+Gaia two-body solution and RV+Hip+GDR2+GDR3. 

\section{Eccentricity Distributions of Brown Dwarfs with Different Semi-major Axis Cuts}

\begin{figure*}[htb]
    \centering
    \includegraphics[width=1.0 \linewidth]{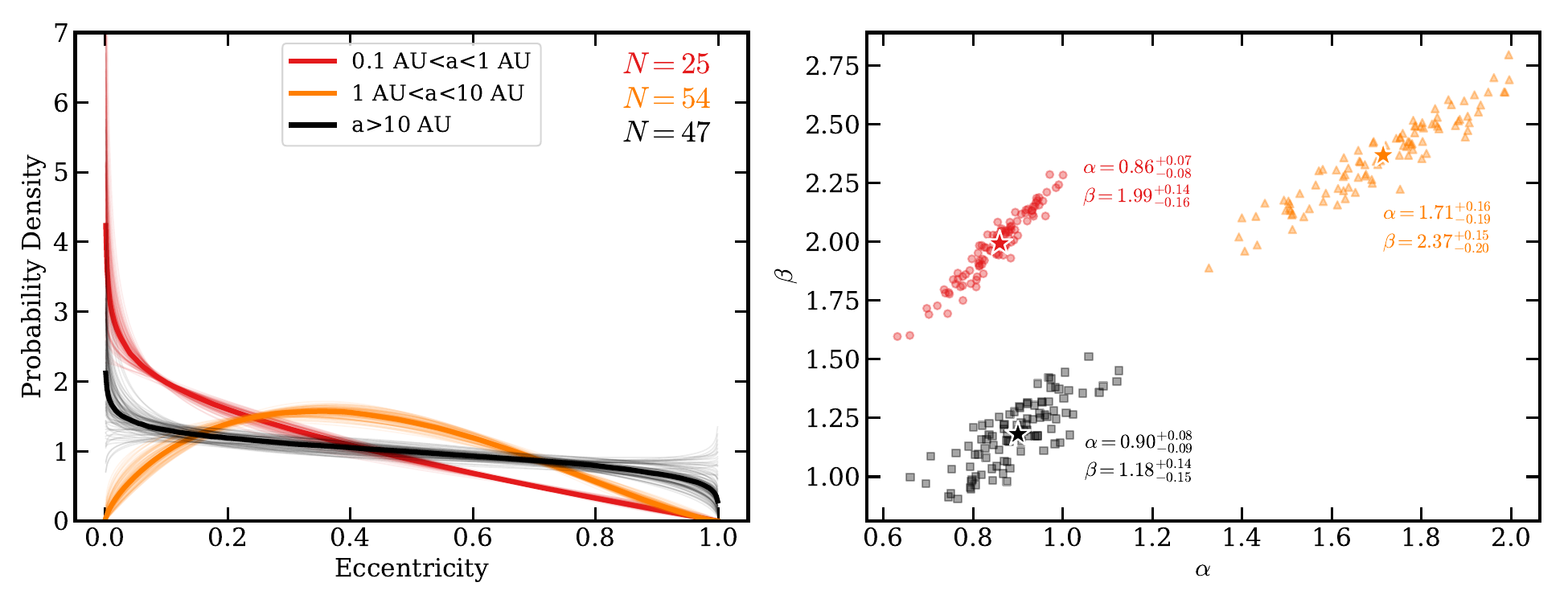}
    \includegraphics[width=1.0 \linewidth]{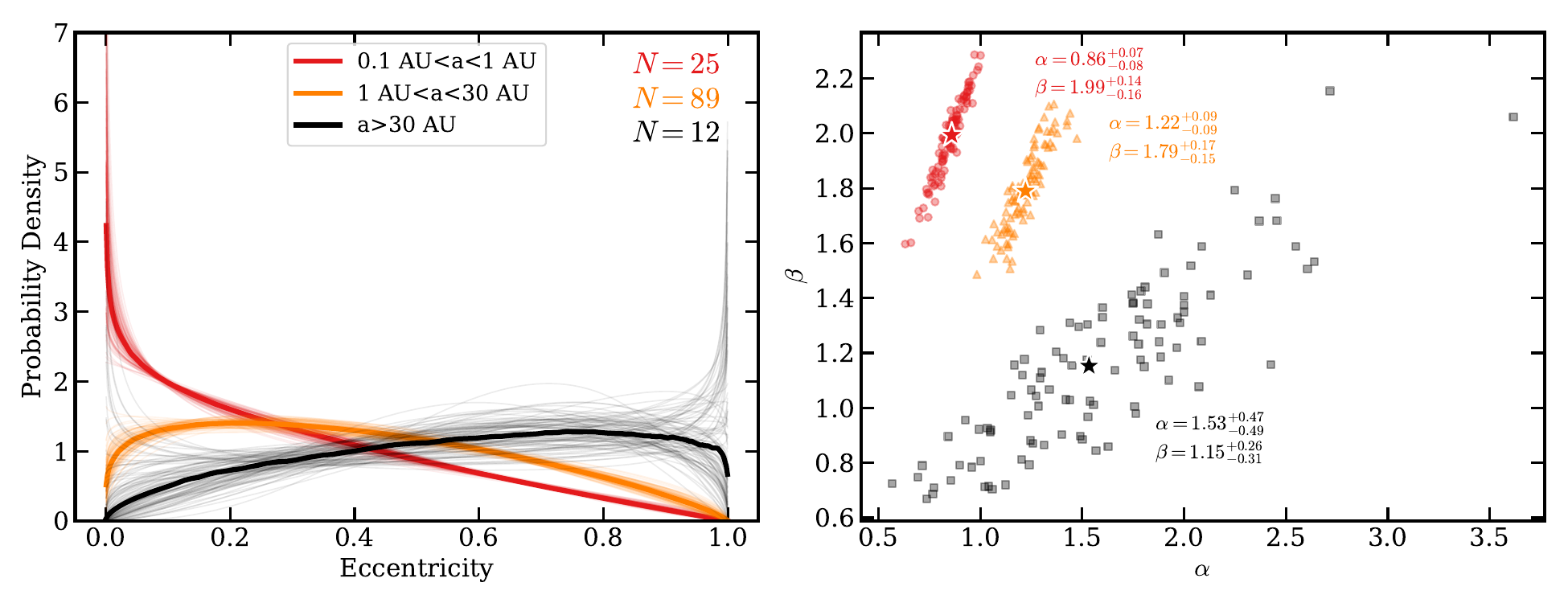}
    \caption{Same as Figure \ref{fig:eccdist}, but using different wide-orbit cutoff of 10 AU (upper panel) and 30 AU (lower panel) to divide the brown dwarf sample into the semi-major axis ranges of 0.1–1 AU, 1–10 AU, >10 AU and 0.1–1 AU, 1–30 AU, >30 AU respectively.The size of each subsample is shown on the upper right corner of each left panel}
    \label{fig:e10AU}
\end{figure*}

Figure~\ref{fig:e10AU} shows the brown dwarf eccentricity distributions with different semi-major axis cuts. 


\bibliography{planet}{}

@ARTICLE{Sagear2026,
       author = {{Sagear}, Sheila and {Ballard}, Sarah and {Gilbert}, Gregory J. and {Albornoz}, Mariangel and {Lam}, Christopher},
        title = "{The Orbital Eccentricity─Radius Relation for Planets Orbiting M Dwarfs}",
      journal = {\aj},
         year = 2026,
        month = feb,
       volume = {171},
       number = {2},
          eid = {115},
        pages = {115},
          doi = {10.3847/1538-3881/ae2d1b},
archivePrefix = {arXiv},
       eprint = {2507.07169},
 primaryClass = {astro-ph.EP},
       adsurl = {https://ui.adsabs.harvard.edu/abs/2026AJ....171..115S}
}

@ARTICLE{Maldonado2017,
       author = {{Maldonado}, J. and {Villaver}, E.},
        title = "{Searching for chemical signatures of brown dwarf formation}",
      journal = {\aap},
         year = 2017,
        month = jun,
       volume = {602},
          eid = {A38},
        pages = {A38},
          doi = {10.1051/0004-6361/201630120},
archivePrefix = {arXiv},
       eprint = {1702.02904},
 primaryClass = {astro-ph.SR},
       adsurl = {https://ui.adsabs.harvard.edu/abs/2017A&A...602A..38M}
}

@ARTICLE{Schlaufman2018,
       author = {{Schlaufman}, Kevin C.},
        title = "{Evidence of an Upper Bound on the Masses of Planets and Its Implications for Giant Planet Formation}",
      journal = {\apj},
         year = 2018,
        month = jan,
       volume = {853},
       number = {1},
          eid = {37},
        pages = {37},
          doi = {10.3847/1538-4357/aa961c},
archivePrefix = {arXiv},
       eprint = {1801.06185},
 primaryClass = {astro-ph.EP},
       adsurl = {https://ui.adsabs.harvard.edu/abs/2018ApJ...853...37S}
}

@ARTICLE{Lindegren:2021,
       author = {{Lindegren}, L. and {Klioner}, S.~A. and {Hern{\'a}ndez}, J. and {Bombrun}, A. and {Ramos-Lerate}, M. and {Steidelm{\"u}ller}, H. and {Bastian}, U. and {Biermann}, M. and {de Torres}, A. and {Gerlach}, E. and {Geyer}, R. and {Hilger}, T. and {Hobbs}, D. and {Lammers}, U. and {McMillan}, P.~J. and {Stephenson}, C.~A. and {Casta{\~n}eda}, J. and {Davidson}, M. and {Fabricius}, C. and {Gracia-Abril}, G. and {Portell}, J. and {Rowell}, N. and {Teyssier}, D. and {Torra}, F. and {Bartolom{\'e}}, S. and {Clotet}, M. and {Garralda}, N. and {Gonz{\'a}lez-Vidal}, J.~J. and {Torra}, J. and {Abbas}, U. and {Altmann}, M. and {Anglada Varela}, E. and {Balaguer-N{\'u}{\~n}ez}, L. and {Balog}, Z. and {Barache}, C. and {Becciani}, U. and {Bernet}, M. and {Bertone}, S. and {Bianchi}, L. and {Bouquillon}, S. and {Brown}, A.~G.~A. and {Bucciarelli}, B. and {Busonero}, D. and {Butkevich}, A.~G. and {Buzzi}, R. and {Cancelliere}, R. and {Carlucci}, T. and {Charlot}, P. and {Cioni}, M.-R.~L. and {Crosta}, M. and {Crowley}, C. and {del Peloso}, E.~F. and {del Pozo}, E. and {Drimmel}, R. and {Esquej}, P. and {Fienga}, A. and {Fraile}, E. and {Gai}, M. and {Garcia-Reinaldos}, M. and {Guerra}, R. and {Hambly}, N.~C. and {Hauser}, M. and {Jan{\ss}en}, K. and {Jordan}, S. and {Kostrzewa-Rutkowska}, Z. and {Lattanzi}, M.~G. and {Liao}, S. and {Licata}, E. and {Lister}, T.~A. and {L{\"o}ffler}, W. and {Marchant}, J.~M. and {Masip}, A. and {Mignard}, F. and {Mints}, A. and {Molina}, D. and {Mora}, A. and {Morbidelli}, R. and {Murphy}, C.~P. and {Pagani}, C. and {Panuzzo}, P. and {Pe{\~n}alosa Esteller}, X. and {Poggio}, E. and {Re Fiorentin}, P. and {Riva}, A. and {Sagrist{\`a} Sell{\'e}s}, A. and {Sanchez Gimenez}, V. and {Sarasso}, M. and {Sciacca}, E. and {Siddiqui}, H.~I. and {Smart}, R.~L. and {Souami}, D. and {Spagna}, A. and {Steele}, I.~A. and {Taris}, F. and {Utrilla}, E. and {van Reeven}, W. and {Vecchiato}, A.},
        title = "{Gaia Early Data Release 3. The astrometric solution}",
      journal = {\aap},
         year = 2021,
        month = may,
       volume = {649},
          eid = {A2},
        pages = {A2},
          doi = {10.1051/0004-6361/202039709},
archivePrefix = {arXiv},
       eprint = {2012.03380},
 primaryClass = {astro-ph.IM},
       adsurl = {https://ui.adsabs.harvard.edu/abs/2021A&A...649A...2L}
}

@ARTICLE{Blunt2026,
       author = {{Blunt}, Sarah and {Wang}, Jason and {Murray-Clay}, Ruth and {Macintosh}, Bruce and {Rubenzahl}, Ryan A. and {Fulton}, B.~J.},
        title = "{Evidence for a Peak at {\ensuremath{\sim}}0.3 in the Eccentricity Distribution of Typical Super-Jovian Exoplanets}",
      journal = {\apjl},
         year = 2026,
        month = mar,
       volume = {999},
       number = {2},
          eid = {L26},
        pages = {L26},
          doi = {10.3847/2041-8213/ae41c3},
archivePrefix = {arXiv},
       eprint = {2601.18877},
 primaryClass = {astro-ph.EP},
       adsurl = {https://ui.adsabs.harvard.edu/abs/2026ApJ...999L..26B}
}

@ARTICLE{mamajek2013ApJS,
       author = {{Pecaut}, Mark J. and {Mamajek}, Eric E.},
        title = "{Intrinsic Colors, Temperatures, and Bolometric Corrections of Pre-main-sequence Stars}",
      journal = {\apjs},
         year = 2013,
        month = sep,
       volume = {208},
       number = {1},
          eid = {9},
        pages = {9},
          doi = {10.1088/0067-0049/208/1/9},
archivePrefix = {arXiv},
       eprint = {1307.2657},
 primaryClass = {astro-ph.SR},
       adsurl = {https://ui.adsabs.harvard.edu/abs/2013ApJS..208....9P}
}

@ARTICLE{HARPSRVBank2,
       author = {{Perdelwitz}, V. and {Trifonov}, T. and {Teklu}, J.~T. and {Sreenivas}, K.~R. and {Tal-Or}, L.},
        title = "{Analysis of the public HARPS/ESO spectroscopic archive. Ca II H\&K time series for the HARPS radial velocity database}",
      journal = {\aap},
         year = 2024,
        month = mar,
       volume = {683},
          eid = {A125},
        pages = {A125},
          doi = {10.1051/0004-6361/202348263},
archivePrefix = {arXiv},
       eprint = {2311.12438},
 primaryClass = {astro-ph.SR},
       adsurl = {https://ui.adsabs.harvard.edu/abs/2024A&A...683A.125P}
}

@ARTICLE{Bitsch2010,
       author = {{Bitsch}, B. and {Kley}, W.},
        title = "{Orbital evolution of eccentric planets in radiative discs}",
      journal = {\aap},
         year = 2010,
        month = nov,
       volume = {523},
          eid = {A30},
        pages = {A30},
          doi = {10.1051/0004-6361/201014414},
archivePrefix = {arXiv},
       eprint = {1008.2656},
 primaryClass = {astro-ph.EP},
       adsurl = {https://ui.adsabs.harvard.edu/abs/2010A&A...523A..30B}
}

@ARTICLE{Cresswell2008,
       author = {{Cresswell}, P. and {Nelson}, R.~P.},
        title = "{Three-dimensional simulations of multiple protoplanets embedded in a protostellar disc}",
      journal = {\aap},
         year = 2008,
        month = may,
       volume = {482},
       number = {2},
        pages = {677-690},
          doi = {10.1051/0004-6361:20079178},
archivePrefix = {arXiv},
       eprint = {0811.4322},
 primaryClass = {astro-ph},
       adsurl = {https://ui.adsabs.harvard.edu/abs/2008A&A...482..677C}
}

@ARTICLE{Cresswell2006,
       author = {{Cresswell}, P. and {Nelson}, R.~P.},
        title = "{On the evolution of multiple protoplanets embedded in a protostellar disc}",
      journal = {\aap},
         year = 2006,
        month = may,
       volume = {450},
       number = {2},
        pages = {833-853},
          doi = {10.1051/0004-6361:20054551},
       adsurl = {https://ui.adsabs.harvard.edu/abs/2006A&A...450..833C}
}

@ARTICLE{Papaloizou2000,
       author = {{Papaloizou}, J.~C.~B. and {Larwood}, J.~D.},
        title = "{On the orbital evolution and growth of protoplanets embedded in a gaseous disc}",
      journal = {\mnras},
         year = 2000,
        month = jul,
       volume = {315},
       number = {4},
        pages = {823-833},
          doi = {10.1046/j.1365-8711.2000.03466.x},
archivePrefix = {arXiv},
       eprint = {astro-ph/9911431},
 primaryClass = {astro-ph},
       adsurl = {https://ui.adsabs.harvard.edu/abs/2000MNRAS.315..823P}
}

@ARTICLE{Ida2020,
       author = {{Ida}, Shigeru and {Muto}, Takayuki and {Matsumura}, Soko and {Brasser}, Ramon},
        title = "{A new and simple prescription for planet orbital migration and eccentricity damping by planet-disc interactions based on dynamical friction}",
      journal = {\mnras},
         year = 2020,
        month = jun,
       volume = {494},
       number = {4},
        pages = {5666-5674},
          doi = {10.1093/mnras/staa1073},
archivePrefix = {arXiv},
       eprint = {2004.07481},
 primaryClass = {astro-ph.EP},
       adsurl = {https://ui.adsabs.harvard.edu/abs/2020MNRAS.494.5666I}
}

@ARTICLE{Barbato2026,
       author = {{Barbato}, D. and {Pinamonti}, M. and {Sozzetti}, A. and {Desidera}, S. and {D'Orazi}, V. and {Maldonado}, J. and {Biazzo}, K. and {Naponiello}, L. and {Lanza}, A.~F. and {Bignamini}, A. and {Bonomo}, A.~S. and {Brogi}, M. and {Cabona}, L. and {Damasso}, M. and {Gratton}, R. and {Mancini}, L. and {Mantovan}, G. and {Nardiello}, D. and {Rainer}, M. and {Guilluy}, G. and {Giacobbe}, P. and {Malavolta}, L. and {Cosentino}, R. and {Boschin}, W. and {Claudi}, R.},
        title = "{The GAPS programme at TNG: LXXV. Validating and confirming Gaia sub-stellar astrometric candidates with HARPS-N}",
      journal = {\aap},
         year = 2026,
        month = aug,
       volume = {712},
          eid = {A213},
        pages = {A213},
          doi = {10.1051/0004-6361/202557585},
archivePrefix = {arXiv},
       eprint = {2606.29001},
 primaryClass = {astro-ph.EP},
       adsurl = {https://ui.adsabs.harvard.edu/abs/2026A&A...712A.213B}
}

@ARTICLE{Kley2012,
       author = {{Kley}, W. and {Nelson}, R.~P.},
        title = "{Planet-Disk Interaction and Orbital Evolution}",
      journal = {\araa},
         year = 2012,
        month = sep,
       volume = {50},
        pages = {211-249},
          doi = {10.1146/annurev-astro-081811-125523},
archivePrefix = {arXiv},
       eprint = {1203.1184},
 primaryClass = {astro-ph.EP},
       adsurl = {https://ui.adsabs.harvard.edu/abs/2012ARA&A..50..211K}
}

@ARTICLE{Jiang2019,
       author = {{Jiang}, Jonathan H. and {Ji}, Xuan and {Cowan}, Nicolas and {Hu}, Renyu and {Zhu}, Zonghong},
        title = "{Empirical Predictions for the Period Distribution of Planets to Be Discovered by the Transiting Exoplanet Survey Satellite}",
      journal = {\aj},
         year = 2019,
        month = aug,
       volume = {158},
       number = {2},
          eid = {96},
        pages = {96},
          doi = {10.3847/1538-3881/ab2a7a},
archivePrefix = {arXiv},
       eprint = {1906.06795},
 primaryClass = {astro-ph.EP},
       adsurl = {https://ui.adsabs.harvard.edu/abs/2019AJ....158...96J}
}

@ARTICLE{Stamatellos2009,
       author = {{Stamatellos}, Dimitris and {Whitworth}, Anthony P.},
        title = "{The properties of brown dwarfs and low-mass hydrogen-burning stars formed by disc fragmentation}",
      journal = {\mnras},
         year = 2009,
        month = jan,
       volume = {392},
       number = {1},
        pages = {413-427},
          doi = {10.1111/j.1365-2966.2008.14069.x},
archivePrefix = {arXiv},
       eprint = {0810.1687},
 primaryClass = {astro-ph},
       adsurl = {https://ui.adsabs.harvard.edu/abs/2009MNRAS.392..413S}
}

@ARTICLE{Hall2017,
       author = {{Hall}, Cassandra and {Forgan}, Duncan and {Rice}, Ken},
        title = "{Identifying and analysing protostellar disc fragments in smoothed particle hydrodynamics simulations}",
      journal = {\mnras},
         year = 2017,
        month = sep,
       volume = {470},
       number = {3},
        pages = {2517-2538},
          doi = {10.1093/mnras/stx1244},
archivePrefix = {arXiv},
       eprint = {1705.06690},
 primaryClass = {astro-ph.EP},
       adsurl = {https://ui.adsabs.harvard.edu/abs/2017MNRAS.470.2517H}
}

@ARTICLE{Boss2000,
       author = {{Boss}, Alan P.},
        title = "{Possible Rapid Gas Giant Planet Formation in the Solar Nebula and Other Protoplanetary Disks}",
      journal = {\apjl},
         year = 2000,
        month = jun,
       volume = {536},
       number = {2},
        pages = {L101-L104},
          doi = {10.1086/312737},
       adsurl = {https://ui.adsabs.harvard.edu/abs/2000ApJ...536L.101B}
}

@BOOK{Cutri2003,
   author = {{Cutri}, R.~M. and {Skrutskie}, M.~F. and {van Dyk}, S. and 
	{Beichman}, C.~A. and {Carpenter}, J.~M. and {Chester}, T. and 
	{Cambresy}, L. and {Evans}, T. and {Fowler}, J. and {Gizis}, J. and 
	{Howard}, E. and {Huchra}, J. and {Jarrett}, T. and {Kopan}, E.~L. and 
	{Kirkpatrick}, J.~D. and {Light}, R.~M. and {Marsh}, K.~A. and 
	{McCallon}, H. and {Schneider}, S. and {Stiening}, R. and {Sykes}, M. and 
	{Weinberg}, M. and {Wheaton}, W.~A. and {Wheelock}, S. and {Zacarias}, N.
	},
    title = "{2MASS All Sky Catalog of point sources.}",
booktitle = {``The IRSA 2MASS All-Sky Point Source Catalog, NASA/IPAC Infrared Science Archive.~<A href=''http://irsa.ipac.caltech.edu/applications/Gator/``>http://irsa.ipac.caltech.edu/applications/Gator/</A>''},
     year = 2003,
    month = jun,
   adsurl = {http://adsabs.harvard.edu/abs/2003tmc..book.....C}
}

@ARTICLE{Foreman2013,
   author = {{Foreman-Mackey}, D. and {Hogg}, D.~W. and {Lang}, D. and {Goodman}, J.
	},
    title = "{emcee: The MCMC Hammer}",
  journal = {\pasp},
archivePrefix = "arXiv",
   eprint = {1202.3665},
 primaryClass = "astro-ph.IM",
     year = 2013,
    month = mar,
   volume = 125,
    pages = {306},
      doi = {10.1086/670067},
   adsurl = {http://adsabs.harvard.edu/abs/2013PASP..125..306F}
}

@ARTICLE{Foreman2016,
       author = {{Foreman-Mackey}, Daniel},
        title = "{corner.py: Scatterplot matrices in Python}",
      journal = {The Journal of Open Source Software},
         year = 2016,
        month = jun,
       volume = {1},
        pages = {24},
          doi = {10.21105/joss.00024},
       adsurl = {https://ui.adsabs.harvard.edu/abs/2016JOSS....1...24F}
}

@ARTICLE{Fulton2018,
       author = {{Fulton}, Benjamin J. and {Petigura}, Erik A. and {Blunt}, Sarah and
         {Sinukoff}, Evan},
        title = "{RadVel: The Radial Velocity Modeling Toolkit}",
      journal = {\pasp},
         year = "2018",
        month = "Apr",
       volume = {130},
       number = {986},
        pages = {044504},
          doi = {10.1088/1538-3873/aaaaa8},
archivePrefix = {arXiv},
       eprint = {1801.01947},
 primaryClass = {astro-ph.IM},
       adsurl = {https://ui.adsabs.harvard.edu/abs/2018PASP..130d4504F}
}

@ARTICLE{hog2000,
       author = {{H{\o}g}, E. and {Fabricius}, C. and {Makarov}, V.~V. and {Urban}, S. and
         {Corbin}, T. and {Wycoff}, G. and {Bastian}, U. and {Schwekendiek}, P. and
         {Wicenec}, A.},
        title = "{The Tycho-2 catalogue of the 2.5 million brightest stars}",
      journal = {\aap},
         year = "2000",
        month = "Mar",
       volume = {355},
        pages = {L27-L30},
       adsurl = {https://ui.adsabs.harvard.edu/abs/2000A&A...355L..27H}
}

@ARTICLE{Kipping2013,
       author = {{Kipping}, David M.},
        title = "{Efficient, uninformative sampling of limb darkening coefficients for two-parameter laws}",
      journal = {\mnras},
         year = "2013",
        month = "Nov",
       volume = {435},
       number = {3},
        pages = {2152-2160},
          doi = {10.1093/mnras/stt1435},
archivePrefix = {arXiv},
       eprint = {1308.0009},
 primaryClass = {astro-ph.SR},
       adsurl = {https://ui.adsabs.harvard.edu/abs/2013MNRAS.435.2152K}
}

@ARTICLE{Gaia2018,
       author = {{Gaia Collaboration} and {Brown}, A.~G.~A. and {Vallenari}, A. and {Prusti}, T. and {de Bruijne}, J.~H.~J. and {Babusiaux}, C. and {Bailer-Jones}, C.~A.~L. and {Biermann}, M. and {Evans}, D.~W. and {Eyer}, L. and {Jansen}, F. and {Jordi}, C. and {Klioner}, S.~A. and {Lammers}, U. and {Lindegren}, L. and {Luri}, X. and {Mignard}, F. and {Panem}, C. and {Pourbaix}, D. and {Randich}, S. and {Sartoretti}, P. and {Siddiqui}, H.~I. and {Soubiran}, C. and {van Leeuwen}, F. and {Walton}, N.~A. and {Arenou}, F. and {Bastian}, U. and {Cropper}, M. and {Drimmel}, R. and {Katz}, D. and {Lattanzi}, M.~G. and {Bakker}, J. and {Cacciari}, C. and {Casta{\~n}eda}, J. and {Chaoul}, L. and {Cheek}, N. and {De Angeli}, F. and {Fabricius}, C. and {Guerra}, R. and {Holl}, B. and {Masana}, E. and {Messineo}, R. and {Mowlavi}, N. and {Nienartowicz}, K. and {Panuzzo}, P. and {Portell}, J. and {Riello}, M. and {Seabroke}, G.~M. and {Tanga}, P. and {Th{\'e}venin}, F. and {Gracia-Abril}, G. and {Comoretto}, G. and {Garcia-Reinaldos}, M. and {Teyssier}, D. and {Altmann}, M. and {Andrae}, R. and {Audard}, M. and {Bellas-Velidis}, I. and {Benson}, K. and {Berthier}, J. and {Blomme}, R. and {Burgess}, P. and {Busso}, G. and {Carry}, B. and {Cellino}, A. and {Clementini}, G. and {Clotet}, M. and {Creevey}, O. and {Davidson}, M. and {De Ridder}, J. and {Delchambre}, L. and {Dell'Oro}, A. and {Ducourant}, C. and {Fern{\'a}ndez-Hern{\'a}ndez}, J. and {Fouesneau}, M. and {Fr{\'e}mat}, Y. and {Galluccio}, L. and {Garc{\'\i}a-Torres}, M. and {Gonz{\'a}lez-N{\'u}{\~n}ez}, J. and {Gonz{\'a}lez-Vidal}, J.~J. and {Gosset}, E. and {Guy}, L.~P. and {Halbwachs}, J. -L. and {Hambly}, N.~C. and {Harrison}, D.~L. and {Hern{\'a}ndez}, J. and {Hestroffer}, D. and {Hodgkin}, S.~T. and {Hutton}, A. and {Jasniewicz}, G. and {Jean-Antoine-Piccolo}, A. and {Jordan}, S. and {Korn}, A.~J. and {Krone-Martins}, A. and {Lanzafame}, A.~C. and {Lebzelter}, T. and {L{\"o}ffler}, W. and {Manteiga}, M. and {Marrese}, P.~M. and {Mart{\'\i}n-Fleitas}, J.~M. and {Moitinho}, A. and {Mora}, A. and {Muinonen}, K. and {Osinde}, J. and {Pancino}, E. and {Pauwels}, T. and {Petit}, J. -M. and {Recio-Blanco}, A. and {Richards}, P.~J. and {Rimoldini}, L. and {Robin}, A.~C. and {Sarro}, L.~M. and {Siopis}, C. and {Smith}, M. and {Sozzetti}, A. and {S{\"u}veges}, M. and {Torra}, J. and {van Reeven}, W. and {Abbas}, U. and {Abreu Aramburu}, A. and {Accart}, S. and {Aerts}, C. and {Altavilla}, G. and {{\'A}lvarez}, M.~A. and {Alvarez}, R. and {Alves}, J. and {Anderson}, R.~I. and {Andrei}, A.~H. and {Anglada Varela}, E. and {Antiche}, E. and {Antoja}, T. and {Arcay}, B. and {Astraatmadja}, T.~L. and {Bach}, N. and {Baker}, S.~G. and {Balaguer-N{\'u}{\~n}ez}, L. and {Balm}, P. and {Barache}, C. and {Barata}, C. and {Barbato}, D. and {Barblan}, F. and {Barklem}, P.~S. and {Barrado}, D. and {Barros}, M. and {Barstow}, M.~A. and {Bartholom{\'e} Mu{\~n}oz}, S. and {Bassilana}, J. -L. and {Becciani}, U. and {Bellazzini}, M. and {Berihuete}, A. and {Bertone}, S. and {Bianchi}, L. and {Bienaym{\'e}}, O. and {Blanco-Cuaresma}, S. and {Boch}, T. and {Boeche}, C. and {Bombrun}, A. and {Borrachero}, R. and {Bossini}, D. and {Bouquillon}, S. and {Bourda}, G. and {Bragaglia}, A. and {Bramante}, L. and {Breddels}, M.~A. and {Bressan}, A. and {Brouillet}, N. and {Br{\"u}semeister}, T. and {Brugaletta}, E. and {Bucciarelli}, B. and {Burlacu}, A. and {Busonero}, D. and {Butkevich}, A.~G. and {Buzzi}, R. and {Caffau}, E. and {Cancelliere}, R. and {Cannizzaro}, G. and {Cantat-Gaudin}, T. and {Carballo}, R. and {Carlucci}, T. and {Carrasco}, J.~M. and {Casamiquela}, L. and {Castellani}, M. and {Castro-Ginard}, A. and {Charlot}, P. and {Chemin}, L. and {Chiavassa}, A. and {Cocozza}, G. and {Costigan}, G. and {Cowell}, S. and {Crifo}, F. and {Crosta}, M. and {Crowley}, C. and {Cuypers}, J. and {Dafonte}, C. and {Damerdji}, Y. and {Dapergolas}, A. and {David}, P. and {David}, M. and {de Laverny}, P. and {De Luise}, F. and {De March}, R. and {de Martino}, D. and {de Souza}, R. and {de Torres}, A. and {Debosscher}, J. and {del Pozo}, E. and {Delbo}, M. and {Delgado}, A. and {Delgado}, H.~E. and {Di Matteo}, P. and {Diakite}, S. and {Diener}, C. and {Distefano}, E. and {Dolding}, C. and {Drazinos}, P. and {Dur{\'a}n}, J. and {Edvardsson}, B. and {Enke}, H. and {Eriksson}, K. and {Esquej}, P. and {Eynard Bontemps}, G. and {Fabre}, C. and {Fabrizio}, M. and {Faigler}, S. and {Falc{\~a}o}, A.~J. and {Farr{\`a}s Casas}, M. and {Federici}, L. and {Fedorets}, G. and {Fernique}, P. and {Figueras}, F. and {Filippi}, F. and {Findeisen}, K. and {Fonti}, A. and {Fraile}, E. and {Fraser}, M. and {Fr{\'e}zouls}, B. and {Gai}, M. and {Galleti}, S. and {Garabato}, D. and {Garc{\'\i}a-Sedano}, F. and {Garofalo}, A. and {Garralda}, N. and {Gavel}, A. and {Gavras}, P. and {Gerssen}, J. and {Geyer}, R. and {Giacobbe}, P. and {Gilmore}, G. and {Girona}, S. and {Giuffrida}, G. and {Glass}, F. and {Gomes}, M. and {Granvik}, M. and {Gueguen}, A. and {Guerrier}, A. and {Guiraud}, J. and {Guti{\'e}rrez-S{\'a}nchez}, R. and {Haigron}, R. and {Hatzidimitriou}, D. and {Hauser}, M. and {Haywood}, M. and {Heiter}, U. and {Helmi}, A. and {Heu}, J. and {Hilger}, T. and {Hobbs}, D. and {Hofmann}, W. and {Holland}, G. and {Huckle}, H.~E. and {Hypki}, A. and {Icardi}, V. and {Jan{\ss}en}, K. and {Jevardat de Fombelle}, G. and {Jonker}, P.~G. and {Juh{\'a}sz}, {\'A}. L. and {Julbe}, F. and {Karampelas}, A. and {Kewley}, A. and {Klar}, J. and {Kochoska}, A. and {Kohley}, R. and {Kolenberg}, K. and {Kontizas}, M. and {Kontizas}, E. and {Koposov}, S.~E. and {Kordopatis}, G. and {Kostrzewa-Rutkowska}, Z. and {Koubsky}, P. and {Lambert}, S. and {Lanza}, A.~F. and {Lasne}, Y. and {Lavigne}, J. -B. and {Le Fustec}, Y. and {Le Poncin-Lafitte}, C. and {Lebreton}, Y. and {Leccia}, S. and {Leclerc}, N. and {Lecoeur-Taibi}, I. and {Lenhardt}, H. and {Leroux}, F. and {Liao}, S. and {Licata}, E. and {Lindstr{\o}m}, H.~E.~P. and {Lister}, T.~A. and {Livanou}, E. and {Lobel}, A. and {L{\'o}pez}, M. and {Managau}, S. and {Mann}, R.~G. and {Mantelet}, G. and {Marchal}, O. and {Marchant}, J.~M. and {Marconi}, M. and {Marinoni}, S. and {Marschalk{\'o}}, G. and {Marshall}, D.~J. and {Martino}, M. and {Marton}, G. and {Mary}, N. and {Massari}, D. and {Matijevi{\v{c}}}, G. and {Mazeh}, T. and {McMillan}, P.~J. and {Messina}, S. and {Michalik}, D. and {Millar}, N.~R. and {Molina}, D. and {Molinaro}, R. and {Moln{\'a}r}, L. and {Montegriffo}, P. and {Mor}, R. and {Morbidelli}, R. and {Morel}, T. and {Morris}, D. and {Mulone}, A.~F. and {Muraveva}, T. and {Musella}, I. and {Nelemans}, G. and {Nicastro}, L. and {Noval}, L. and {O'Mullane}, W. and {Ord{\'e}novic}, C. and {Ord{\'o}{\~n}ez-Blanco}, D. and {Osborne}, P. and {Pagani}, C. and {Pagano}, I. and {Pailler}, F. and {Palacin}, H. and {Palaversa}, L. and {Panahi}, A. and {Pawlak}, M. and {Piersimoni}, A.~M. and {Pineau}, F. -X. and {Plachy}, E. and {Plum}, G. and {Poggio}, E. and {Poujoulet}, E. and {Pr{\v{s}}a}, A. and {Pulone}, L. and {Racero}, E. and {Ragaini}, S. and {Rambaux}, N. and {Ramos-Lerate}, M. and {Regibo}, S. and {Reyl{\'e}}, C. and {Riclet}, F. and {Ripepi}, V. and {Riva}, A. and {Rivard}, A. and {Rixon}, G. and {Roegiers}, T. and {Roelens}, M. and {Romero-G{\'o}mez}, M. and {Rowell}, N. and {Royer}, F. and {Ruiz-Dern}, L. and {Sadowski}, G. and {Sagrist{\`a} Sell{\'e}s}, T. and {Sahlmann}, J. and {Salgado}, J. and {Salguero}, E. and {Sanna}, N. and {Santana-Ros}, T. and {Sarasso}, M. and {Savietto}, H. and {Schultheis}, M. and {Sciacca}, E. and {Segol}, M. and {Segovia}, J.~C. and {S{\'e}gransan}, D. and {Shih}, I. -C. and {Siltala}, L. and {Silva}, A.~F. and {Smart}, R.~L. and {Smith}, K.~W. and {Solano}, E. and {Solitro}, F. and {Sordo}, R. and {Soria Nieto}, S. and {Souchay}, J. and {Spagna}, A. and {Spoto}, F. and {Stampa}, U. and {Steele}, I.~A. and {Steidelm{\"u}ller}, H. and {Stephenson}, C.~A. and {Stoev}, H. and {Suess}, F.~F. and {Surdej}, J. and {Szabados}, L. and {Szegedi-Elek}, E. and {Tapiador}, D. and {Taris}, F. and {Tauran}, G. and {Taylor}, M.~B. and {Teixeira}, R. and {Terrett}, D. and {Teyssandier}, P. and {Thuillot}, W. and {Titarenko}, A. and {Torra Clotet}, F. and {Turon}, C. and {Ulla}, A. and {Utrilla}, E. and {Uzzi}, S. and {Vaillant}, M. and {Valentini}, G. and {Valette}, V. and {van Elteren}, A. and {Van Hemelryck}, E. and {van Leeuwen}, M. and {Vaschetto}, M. and {Vecchiato}, A. and {Veljanoski}, J. and {Viala}, Y. and {Vicente}, D. and {Vogt}, S. and {von Essen}, C. and {Voss}, H. and {Votruba}, V. and {Voutsinas}, S. and {Walmsley}, G. and {Weiler}, M. and {Wertz}, O. and {Wevers}, T. and {Wyrzykowski}, {\L}. and {Yoldas}, A. and {{\v{Z}}erjal}, M. and {Ziaeepour}, H. and {Zorec}, J. and {Zschocke}, S. and {Zucker}, S. and {Zurbach}, C. and {Zwitter}, T.},
        title = "{Gaia Data Release 2. Summary of the contents and survey properties}",
      journal = {\aap},
         year = 2018,
        month = aug,
       volume = {616},
          eid = {A1},
        pages = {A1},
          doi = {10.1051/0004-6361/201833051},
archivePrefix = {arXiv},
       eprint = {1804.09365},
 primaryClass = {astro-ph.GA},
       adsurl = {https://ui.adsabs.harvard.edu/abs/2018A&A...616A...1G}
}

@ARTICLE{Gaia2023,
       author = {{Gaia Collaboration} and {Vallenari}, A. and {Brown}, A.~G.~A. and {Prusti}, T. and {de Bruijne}, J.~H.~J. and {Arenou}, F. and {Babusiaux}, C. and {Biermann}, M. and {Creevey}, O.~L. and {Ducourant}, C. and {Evans}, D.~W. and {Eyer}, L. and {Guerra}, R. and {Hutton}, A. and {Jordi}, C. and {Klioner}, S.~A. and {Lammers}, U.~L. and {Lindegren}, L. and {Luri}, X. and {Mignard}, F. and {Panem}, C. and {Pourbaix}, D. and {Randich}, S. and {Sartoretti}, P. and {Soubiran}, C. and {Tanga}, P. and {Walton}, N.~A. and {Bailer-Jones}, C.~A.~L. and {Bastian}, U. and {Drimmel}, R. and {Jansen}, F. and {Katz}, D. and {Lattanzi}, M.~G. and {van Leeuwen}, F. and {Bakker}, J. and {Cacciari}, C. and {Casta{\~n}eda}, J. and {De Angeli}, F. and {Fabricius}, C. and {Fouesneau}, M. and {Fr{\'e}mat}, Y. and {Galluccio}, L. and {Guerrier}, A. and {Heiter}, U. and {Masana}, E. and {Messineo}, R. and {Mowlavi}, N. and {Nicolas}, C. and {Nienartowicz}, K. and {Pailler}, F. and {Panuzzo}, P. and {Riclet}, F. and {Roux}, W. and {Seabroke}, G.~M. and {Sordo}, R. and {Th{\'e}venin}, F. and {Gracia-Abril}, G. and {Portell}, J. and {Teyssier}, D. and {Altmann}, M. and {Andrae}, R. and {Audard}, M. and {Bellas-Velidis}, I. and {Benson}, K. and {Berthier}, J. and {Blomme}, R. and {Burgess}, P.~W. and {Busonero}, D. and {Busso}, G. and {C{\'a}novas}, H. and {Carry}, B. and {Cellino}, A. and {Cheek}, N. and {Clementini}, G. and {Damerdji}, Y. and {Davidson}, M. and {de Teodoro}, P. and {Nu{\~n}ez Campos}, M. and {Delchambre}, L. and {Dell'Oro}, A. and {Esquej}, P. and {Fern{\'a}ndez-Hern{\'a}ndez}, J. and {Fraile}, E. and {Garabato}, D. and {Garc{\'\i}a-Lario}, P. and {Gosset}, E. and {Haigron}, R. and {Halbwachs}, J. -L. and {Hambly}, N.~C. and {Harrison}, D.~L. and {Hern{\'a}ndez}, J. and {Hestroffer}, D. and {Hodgkin}, S.~T. and {Holl}, B. and {Jan{\ss}en}, K. and {Jevardat de Fombelle}, G. and {Jordan}, S. and {Krone-Martins}, A. and {Lanzafame}, A.~C. and {L{\"o}ffler}, W. and {Marchal}, O. and {Marrese}, P.~M. and {Moitinho}, A. and {Muinonen}, K. and {Osborne}, P. and {Pancino}, E. and {Pauwels}, T. and {Recio-Blanco}, A. and {Reyl{\'e}}, C. and {Riello}, M. and {Rimoldini}, L. and {Roegiers}, T. and {Rybizki}, J. and {Sarro}, L.~M. and {Siopis}, C. and {Smith}, M. and {Sozzetti}, A. and {Utrilla}, E. and {van Leeuwen}, M. and {Abbas}, U. and {{\'A}brah{\'a}m}, P. and {Abreu Aramburu}, A. and {Aerts}, C. and {Aguado}, J.~J. and {Ajaj}, M. and {Aldea-Montero}, F. and {Altavilla}, G. and {{\'A}lvarez}, M.~A. and {Alves}, J. and {Anders}, F. and {Anderson}, R.~I. and {Anglada Varela}, E. and {Antoja}, T. and {Baines}, D. and {Baker}, S.~G. and {Balaguer-N{\'u}{\~n}ez}, L. and {Balbinot}, E. and {Balog}, Z. and {Barache}, C. and {Barbato}, D. and {Barros}, M. and {Barstow}, M.~A. and {Bartolom{\'e}}, S. and {Bassilana}, J. -L. and {Bauchet}, N. and {Becciani}, U. and {Bellazzini}, M. and {Berihuete}, A. and {Bernet}, M. and {Bertone}, S. and {Bianchi}, L. and {Binnenfeld}, A. and {Blanco-Cuaresma}, S. and {Blazere}, A. and {Boch}, T. and {Bombrun}, A. and {Bossini}, D. and {Bouquillon}, S. and {Bragaglia}, A. and {Bramante}, L. and {Breedt}, E. and {Bressan}, A. and {Brouillet}, N. and {Brugaletta}, E. and {Bucciarelli}, B. and {Burlacu}, A. and {Butkevich}, A.~G. and {Buzzi}, R. and {Caffau}, E. and {Cancelliere}, R. and {Cantat-Gaudin}, T. and {Carballo}, R. and {Carlucci}, T. and {Carnerero}, M.~I. and {Carrasco}, J.~M. and {Casamiquela}, L. and {Castellani}, M. and {Castro-Ginard}, A. and {Chaoul}, L. and {Charlot}, P. and {Chemin}, L. and {Chiaramida}, V. and {Chiavassa}, A. and {Chornay}, N. and {Comoretto}, G. and {Contursi}, G. and {Cooper}, W.~J. and {Cornez}, T. and {Cowell}, S. and {Crifo}, F. and {Cropper}, M. and {Crosta}, M. and {Crowley}, C. and {Dafonte}, C. and {Dapergolas}, A. and {David}, M. and {David}, P. and {de Laverny}, P. and {De Luise}, F. and {De March}, R. and {De Ridder}, J. and {de Souza}, R. and {de Torres}, A. and {del Peloso}, E.~F. and {del Pozo}, E. and {Delbo}, M. and {Delgado}, A. and {Delisle}, J. -B. and {Demouchy}, C. and {Dharmawardena}, T.~E. and {Di Matteo}, P. and {Diakite}, S. and {Diener}, C. and {Distefano}, E. and {Dolding}, C. and {Edvardsson}, B. and {Enke}, H. and {Fabre}, C. and {Fabrizio}, M. and {Faigler}, S. and {Fedorets}, G. and {Fernique}, P. and {Fienga}, A. and {Figueras}, F. and {Fournier}, Y. and {Fouron}, C. and {Fragkoudi}, F. and {Gai}, M. and {Garcia-Gutierrez}, A. and {Garcia-Reinaldos}, M. and {Garc{\'\i}a-Torres}, M. and {Garofalo}, A. and {Gavel}, A. and {Gavras}, P. and {Gerlach}, E. and {Geyer}, R. and {Giacobbe}, P. and {Gilmore}, G. and {Girona}, S. and {Giuffrida}, G. and {Gomel}, R. and {Gomez}, A. and {Gonz{\'a}lez-N{\'u}{\~n}ez}, J. and {Gonz{\'a}lez-Santamar{\'\i}a}, I. and {Gonz{\'a}lez-Vidal}, J.~J. and {Granvik}, M. and {Guillout}, P. and {Guiraud}, J. and {Guti{\'e}rrez-S{\'a}nchez}, R. and {Guy}, L.~P. and {Hatzidimitriou}, D. and {Hauser}, M. and {Haywood}, M. and {Helmer}, A. and {Helmi}, A. and {Sarmiento}, M.~H. and {Hidalgo}, S.~L. and {Hilger}, T. and {H{\l}adczuk}, N. and {Hobbs}, D. and {Holland}, G. and {Huckle}, H.~E. and {Jardine}, K. and {Jasniewicz}, G. and {Jean-Antoine Piccolo}, A. and {Jim{\'e}nez-Arranz}, {\'O}. and {Jorissen}, A. and {Juaristi Campillo}, J. and {Julbe}, F. and {Karbevska}, L. and {Kervella}, P. and {Khanna}, S. and {Kontizas}, M. and {Kordopatis}, G. and {Korn}, A.~J. and {K{\'o}sp{\'a}l}, {\'A}. and {Kostrzewa-Rutkowska}, Z. and {Kruszy{\'n}ska}, K. and {Kun}, M. and {Laizeau}, P. and {Lambert}, S. and {Lanza}, A.~F. and {Lasne}, Y. and {Le Campion}, J. -F. and {Lebreton}, Y. and {Lebzelter}, T. and {Leccia}, S. and {Leclerc}, N. and {Lecoeur-Taibi}, I. and {Liao}, S. and {Licata}, E.~L. and {Lindstr{\o}m}, H.~E.~P. and {Lister}, T.~A. and {Livanou}, E. and {Lobel}, A. and {Lorca}, A. and {Loup}, C. and {Madrero Pardo}, P. and {Magdaleno Romeo}, A. and {Managau}, S. and {Mann}, R.~G. and {Manteiga}, M. and {Marchant}, J.~M. and {Marconi}, M. and {Marcos}, J. and {Marcos Santos}, M.~M.~S. and {Mar{\'\i}n Pina}, D. and {Marinoni}, S. and {Marocco}, F. and {Marshall}, D.~J. and {Martin Polo}, L. and {Mart{\'\i}n-Fleitas}, J.~M. and {Marton}, G. and {Mary}, N. and {Masip}, A. and {Massari}, D. and {Mastrobuono-Battisti}, A. and {Mazeh}, T. and {McMillan}, P.~J. and {Messina}, S. and {Michalik}, D. and {Millar}, N.~R. and {Mints}, A. and {Molina}, D. and {Molinaro}, R. and {Moln{\'a}r}, L. and {Monari}, G. and {Mongui{\'o}}, M. and {Montegriffo}, P. and {Montero}, A. and {Mor}, R. and {Mora}, A. and {Morbidelli}, R. and {Morel}, T. and {Morris}, D. and {Muraveva}, T. and {Murphy}, C.~P. and {Musella}, I. and {Nagy}, Z. and {Noval}, L. and {Oca{\~n}a}, F. and {Ogden}, A. and {Ordenovic}, C. and {Osinde}, J.~O. and {Pagani}, C. and {Pagano}, I. and {Palaversa}, L. and {Palicio}, P.~A. and {Pallas-Quintela}, L. and {Panahi}, A. and {Payne-Wardenaar}, S. and {Pe{\~n}alosa Esteller}, X. and {Penttil{\"a}}, A. and {Pichon}, B. and {Piersimoni}, A.~M. and {Pineau}, F. -X. and {Plachy}, E. and {Plum}, G. and {Poggio}, E. and {Pr{\v{s}}a}, A. and {Pulone}, L. and {Racero}, E. and {Ragaini}, S. and {Rainer}, M. and {Raiteri}, C.~M. and {Rambaux}, N. and {Ramos}, P. and {Ramos-Lerate}, M. and {Re Fiorentin}, P. and {Regibo}, S. and {Richards}, P.~J. and {Rios Diaz}, C. and {Ripepi}, V. and {Riva}, A. and {Rix}, H. -W. and {Rixon}, G. and {Robichon}, N. and {Robin}, A.~C. and {Robin}, C. and {Roelens}, M. and {Rogues}, H.~R.~O. and {Rohrbasser}, L. and {Romero-G{\'o}mez}, M. and {Rowell}, N. and {Royer}, F. and {Ruz Mieres}, D. and {Rybicki}, K.~A. and {Sadowski}, G. and {S{\'a}ez N{\'u}{\~n}ez}, A. and {Sagrist{\`a} Sell{\'e}s}, A. and {Sahlmann}, J. and {Salguero}, E. and {Samaras}, N. and {Sanchez Gimenez}, V. and {Sanna}, N. and {Santove{\~n}a}, R. and {Sarasso}, M. and {Schultheis}, M. and {Sciacca}, E. and {Segol}, M. and {Segovia}, J.~C. and {S{\'e}gransan}, D. and {Semeux}, D. and {Shahaf}, S. and {Siddiqui}, H.~I. and {Siebert}, A. and {Siltala}, L. and {Silvelo}, A. and {Slezak}, E. and {Slezak}, I. and {Smart}, R.~L. and {Snaith}, O.~N. and {Solano}, E. and {Solitro}, F. and {Souami}, D. and {Souchay}, J. and {Spagna}, A. and {Spina}, L. and {Spoto}, F. and {Steele}, I.~A. and {Steidelm{\"u}ller}, H. and {Stephenson}, C.~A. and {S{\"u}veges}, M. and {Surdej}, J. and {Szabados}, L. and {Szegedi-Elek}, E. and {Taris}, F. and {Taylor}, M.~B. and {Teixeira}, R. and {Tolomei}, L. and {Tonello}, N. and {Torra}, F. and {Torra}, J. and {Torralba Elipe}, G. and {Trabucchi}, M. and {Tsounis}, A.~T. and {Turon}, C. and {Ulla}, A. and {Unger}, N. and {Vaillant}, M.~V. and {van Dillen}, E. and {van Reeven}, W. and {Vanel}, O. and {Vecchiato}, A. and {Viala}, Y. and {Vicente}, D. and {Voutsinas}, S. and {Weiler}, M. and {Wevers}, T. and {Wyrzykowski}, {\L}. and {Yoldas}, A. and {Yvard}, P. and {Zhao}, H. and {Zorec}, J. and {Zucker}, S. and {Zwitter}, T.},
        title = "{Gaia Data Release 3. Summary of the content and survey properties}",
      journal = {\aap},
         year = 2023,
        month = jun,
       volume = {674},
          eid = {A1},
        pages = {A1},
          doi = {10.1051/0004-6361/202243940},
archivePrefix = {arXiv},
       eprint = {2208.00211},
 primaryClass = {astro-ph.GA},
       adsurl = {https://ui.adsabs.harvard.edu/abs/2023A&A...674A...1G}
}

@ARTICLE{Lo2015,
       author = {{Lo Curto}, G. and {Pepe}, F. and {Avila}, G. and {Boffin}, H. and
         {Bovay}, S. and {Chazelas}, B. and {Coffinet}, A. and {Fleury}, M. and
         {Hughes}, I. and {Lovis}, C. and {Maire}, C. and {Manescau}, A. and
         {Pasquini}, L. and {Rihs}, S. and {Sinclaire}, P. and {Udry}, S.},
        title = "{HARPS Gets New Fibres After 12 Years of Operations}",
      journal = {The Messenger},
         year = 2015,
        month = dec,
       volume = {162},
        pages = {9-15},
       adsurl = {https://ui.adsabs.harvard.edu/abs/2015Msngr.162....9L}
}

@ARTICLE{Ma2014,
       author = {{Ma}, Bo and {Ge}, Jian},
        title = "{Statistical properties of brown dwarf companions: implications for different formation mechanisms}",
      journal = {\mnras},
         year = 2014,
        month = apr,
       volume = {439},
       number = {3},
        pages = {2781-2789},
          doi = {10.1093/mnras/stu134},
archivePrefix = {arXiv},
       eprint = {1303.6442},
 primaryClass = {astro-ph.EP},
       adsurl = {https://ui.adsabs.harvard.edu/abs/2014MNRAS.439.2781M}
}

@ARTICLE{Mayor2003,
   author = {{Mayor}, M. and {Pepe}, F. and {Queloz}, D. and {Bouchy}, F. and 
	{Rupprecht}, G. and {Lo Curto}, G. and {Avila}, G. and {Benz}, W. and 
	{Bertaux}, J.-L. and {Bonfils}, X. and {Dall}, T. and {Dekker}, H. and 
	{Delabre}, B. and {Eckert}, W. and {Fleury}, M. and {Gilliotte}, A. and 
	{Gojak}, D. and {Guzman}, J.~C. and {Kohler}, D. and {Lizon}, J.-L. and 
	{Longinotti}, A. and {Lovis}, C. and {Megevand}, D. and {Pasquini}, L. and 
	{Reyes}, J. and {Sivan}, J.-P. and {Sosnowska}, D. and {Soto}, R. and 
	{Udry}, S. and {van Kesteren}, A. and {Weber}, L. and {Weilenmann}, U.
	},
    title = "{Setting New Standards with HARPS}",
  journal = {The Messenger},
     year = 2003,
    month = dec,
   volume = 114,
    pages = {20-24},
   adsurl = {http://adsabs.harvard.edu/abs/2003Msngr.114...20M}
}

@ARTICLE{Morrissey2007,
       author = {{Morrissey}, Patrick and {Conrow}, Tim and {Barlow}, Tom A. and
         {Small}, Todd and {Seibert}, Mark and {Wyder}, Ted K. and
         {Budav{\'a}ri}, Tam{\'a}s and {Arnouts}, Stephane and
         {Friedman}, Peter G. and {Forster}, Karl and {Martin}, D. Christopher and
         {Neff}, Susan G. and {Schiminovich}, David and {Bianchi}, Luciana and
         {Donas}, Jos{\'e} and {Heckman}, Timothy M. and {Lee}, Young-Wook and
         {Madore}, Barry F. and {Milliard}, Bruno and {Rich}, R. Michael and
         {Szalay}, Alex S. and {Welsh}, Barry Y. and {Yi}, Sukyoung K.},
        title = "{The Calibration and Data Products of GALEX}",
      journal = {\apjs},
         year = 2007,
        month = dec,
       volume = {173},
       number = {2},
        pages = {682-697},
          doi = {10.1086/520512},
       adsurl = {https://ui.adsabs.harvard.edu/abs/2007ApJS..173..682M}
}

@ARTICLE{Pollack1996,
       author = {{Pollack}, James B. and {Hubickyj}, Olenka and {Bodenheimer}, Peter and
         {Lissauer}, Jack J. and {Podolak}, Morris and {Greenzweig}, Yuval},
        title = "{Formation of the Giant Planets by Concurrent Accretion of Solids and Gas}",
      journal = {\icarus},
         year = 1996,
        month = nov,
       volume = {124},
       number = {1},
        pages = {62-85},
          doi = {10.1006/icar.1996.0190},
       adsurl = {https://ui.adsabs.harvard.edu/abs/1996Icar..124...62P}
}

@ARTICLE{Trifonov2020,
       author = {{Trifonov}, Trifon and {Tal-Or}, Lev and {Zechmeister}, Mathias and
         {Kaminski}, Adrian and {Zucker}, Shay and {Mazeh}, Tsevi},
        title = "{Public HARPS radial velocity database corrected for systematic errors}",
      journal = {\aap},
         year = 2020,
        month = apr,
       volume = {636},
          eid = {A74},
        pages = {A74},
          doi = {10.1051/0004-6361/201936686},
archivePrefix = {arXiv},
       eprint = {2001.05942},
 primaryClass = {astro-ph.EP},
       adsurl = {https://ui.adsabs.harvard.edu/abs/2020A&A...636A..74T}
}

@ARTICLE{Speagle2019,
       author = {{Speagle}, Joshua S.},
        title = "{DYNESTY: a dynamic nested sampling package for estimating Bayesian posteriors and evidences}",
      journal = {\mnras},
         year = 2020,
        month = apr,
       volume = {493},
       number = {3},
        pages = {3132-3158},
          doi = {10.1093/mnras/staa278},
archivePrefix = {arXiv},
       eprint = {1904.02180},
 primaryClass = {astro-ph.IM},
       adsurl = {https://ui.adsabs.harvard.edu/abs/2020MNRAS.493.3132S}
}

@ARTICLE{Stassun2019tic,
       author = {{Stassun}, Keivan G. and {Oelkers}, Ryan J. and {Paegert}, Martin and
         {Torres}, Guillermo and {Pepper}, Joshua and {De Lee}, Nathan and
         {Collins}, Kevin and {Latham}, David W. and {Muirhead}, Philip S. and
         {Chittidi}, Jay and {Rojas-Ayala}, B{\'a}rbara and {Fleming}, Scott W. and
         {Rose}, Mark E. and {Tenenbaum}, Peter and {Ting}, Eric B. and
         {Kane}, Stephen R. and {Barclay}, Thomas and {Bean}, Jacob L. and
         {Brassuer}, C.~E. and {Charbonneau}, David and {Ge}, Jian and
         {Lissauer}, Jack J. and {Mann}, Andrew W. and {McLean}, Brian and
         {Mullally}, Susan and {Narita}, Norio and {Plavchan}, Peter and
         {Ricker}, George R. and {Sasselov}, Dimitar and {Seager}, S. and
         {Sharma}, Sanjib and {Shiao}, Bernie and {Sozzetti}, Alessandro and
         {Stello}, Dennis and {Vanderspek}, Roland and {Wallace}, Geoff and
         {Winn}, Joshua N.},
        title = "{The Revised TESS Input Catalog and Candidate Target List}",
      journal = {\aj},
         year = "2019",
        month = "Oct",
       volume = {158},
       number = {4},
          eid = {138},
        pages = {138},
          doi = {10.3847/1538-3881/ab3467},
archivePrefix = {arXiv},
       eprint = {1905.10694},
 primaryClass = {astro-ph.SR},
       adsurl = {https://ui.adsabs.harvard.edu/abs/2019AJ....158..138S}
}

@ARTICLE{Sullivan2015,
       author = {{Sullivan}, Peter W. and {Winn}, Joshua N. and
         {Berta-Thompson}, Zachory K. and {Charbonneau}, David and
         {Deming}, Drake and {Dressing}, Courtney D. and {Latham}, David W. and
         {Levine}, Alan M. and {McCullough}, Peter R. and {Morton}, Timothy and
         {Ricker}, George R. and {Vanderspek}, Roland and {Woods}, Deborah},
        title = "{The Transiting Exoplanet Survey Satellite: Simulations of Planet Detections and Astrophysical False Positives}",
      journal = {\apj},
         year = "2015",
        month = "Aug",
       volume = {809},
       number = {1},
          eid = {77},
        pages = {77},
          doi = {10.1088/0004-637X/809/1/77},
archivePrefix = {arXiv},
       eprint = {1506.03845},
 primaryClass = {astro-ph.EP},
       adsurl = {https://ui.adsabs.harvard.edu/abs/2015ApJ...809...77S}
}

@ARTICLE{wright2010,
       author = {{Wright}, Edward L. and {Eisenhardt}, Peter R.~M. and {Mainzer}, Amy K. and
         {Ressler}, Michael E. and {Cutri}, Roc M. and {Jarrett}, Thomas and
         {Kirkpatrick}, J. Davy and {Padgett}, Deborah and
         {McMillan}, Robert S. and {Skrutskie}, Michael and {Stanford}, S.~A. and
         {Cohen}, Martin and {Walker}, Russell G. and {Mather}, John C. and
         {Leisawitz}, David and {Gautier}, Thomas N., III and {McLean}, Ian and
         {Benford}, Dominic and {Lonsdale}, Carol J. and {Blain}, Andrew and
         {Mendez}, Bryan and {Irace}, William R. and {Duval}, Valerie and
         {Liu}, Fengchuan and {Royer}, Don and {Heinrichsen}, Ingolf and
         {Howard}, Joan and {Shannon}, Mark and {Kendall}, Martha and
         {Walsh}, Amy L. and {Larsen}, Mark and {Cardon}, Joel G. and
         {Schick}, Scott and {Schwalm}, Mark and {Abid}, Mohamed and
         {Fabinsky}, Beth and {Naes}, Larry and {Tsai}, Chao-Wei},
        title = "{The Wide-field Infrared Survey Explorer (WISE): Mission Description and Initial On-orbit Performance}",
      journal = {\aj},
         year = "2010",
        month = "Dec",
       volume = {140},
       number = {6},
        pages = {1868-1881},
          doi = {10.1088/0004-6256/140/6/1868},
archivePrefix = {arXiv},
       eprint = {1008.0031},
 primaryClass = {astro-ph.IM},
       adsurl = {https://ui.adsabs.harvard.edu/abs/2010AJ....140.1868W}
}

@ARTICLE{Yee2017,
       author = {{Yee}, Samuel W. and {Petigura}, Erik A. and {von Braun}, Kaspar},
        title = "{Precision Stellar Characterization of FGKM Stars using an Empirical Spectral Library}",
      journal = {\apj},
         year = "2017",
        month = "Feb",
       volume = {836},
       number = {1},
          eid = {77},
        pages = {77},
          doi = {10.3847/1538-4357/836/1/77},
archivePrefix = {arXiv},
       eprint = {1701.00922},
 primaryClass = {astro-ph.SR},
       adsurl = {https://ui.adsabs.harvard.edu/abs/2017ApJ...836...77Y}
}

@ARTICLE{Zechmeister2018,
       author = {{Zechmeister}, M. and {Reiners}, A. and {Amado}, P.~J. and {Azzaro}, M. and
         {Bauer}, F.~F. and {B{\'e}jar}, V.~J.~S. and {Caballero}, J.~A. and
         {Guenther}, E.~W. and {Hagen}, H. -J. and {Jeffers}, S.~V. and
         {Kaminski}, A. and {K{\"u}rster}, M. and {Launhardt}, R. and
         {Montes}, D. and {Morales}, J.~C. and {Quirrenbach}, A. and
         {Reffert}, S. and {Ribas}, I. and {Seifert}, W. and {Tal-Or}, L. and
         {Wolthoff}, V.},
        title = "{Spectrum radial velocity analyser (SERVAL). High-precision radial velocities and two alternative spectral indicators}",
      journal = {\aap},
         year = "2018",
        month = "Jan",
       volume = {609},
          eid = {A12},
        pages = {A12},
          doi = {10.1051/0004-6361/201731483},
archivePrefix = {arXiv},
       eprint = {1710.10114},
 primaryClass = {astro-ph.IM},
       adsurl = {https://ui.adsabs.harvard.edu/abs/2018A&A...609A..12Z}
}

@ARTICLE{Lindegren:2018,
   author = {{Lindegren}, L. and {Hern{\'a}ndez}, J. and {Bombrun}, A. and 
	{Klioner}, S. and {Bastian}, U. and {Ramos-Lerate}, M. and {de Torres}, A. and 
	{Steidelm{\"u}ller}, H. and {Stephenson}, C. and {Hobbs}, D. and 
	{Lammers}, U. and {Biermann}, M. and {Geyer}, R. and {Hilger}, T. and 
	{Michalik}, D. and {Stampa}, U. and {McMillan}, P.~J. and {Casta{\~n}eda}, J. and 
	{Clotet}, M. and {Comoretto}, G. and {Davidson}, M. and {Fabricius}, C. and 
	{Gracia}, G. and {Hambly}, N.~C. and {Hutton}, A. and {Mora}, A. and 
	{Portell}, J. and {van Leeuwen}, F. and {Abbas}, U. and {Abreu}, A. and 
	{Altmann}, M. and {Andrei}, A. and {Anglada}, E. and {Balaguer-N{\'u}{\~n}ez}, L. and 
	{Barache}, C. and {Becciani}, U. and {Bertone}, S. and {Bianchi}, L. and 
	{Bouquillon}, S. and {Bourda}, G. and {Br{\"u}semeister}, T. and 
	{Bucciarelli}, B. and {Busonero}, D. and {Buzzi}, R. and {Cancelliere}, R. and 
	{Carlucci}, T. and {Charlot}, P. and {Cheek}, N. and {Crosta}, M. and 
	{Crowley}, C. and {de Bruijne}, J. and {de Felice}, F. and {Drimmel}, R. and 
	{Esquej}, P. and {Fienga}, A. and {Fraile}, E. and {Gai}, M. and 
	{Garralda}, N. and {Gonz{\'a}lez-Vidal}, J.~J. and {Guerra}, R. and 
	{Hauser}, M. and {Hofmann}, W. and {Holl}, B. and {Jordan}, S. and 
	{Lattanzi}, M.~G. and {Lenhardt}, H. and {Liao}, S. and {Licata}, E. and 
	{Lister}, T. and {L{\"o}ffler}, W. and {Marchant}, J. and {Martin-Fleitas}, J.-M. and 
	{Messineo}, R. and {Mignard}, F. and {Morbidelli}, R. and {Poggio}, E. and 
	{Riva}, A. and {Rowell}, N. and {Salguero}, E. and {Sarasso}, M. and 
	{Sciacca}, E. and {Siddiqui}, H. and {Smart}, R.~L. and {Spagna}, A. and 
	{Steele}, I. and {Taris}, F. and {Torra}, J. and {van Elteren}, A. and 
	{van Reeven}, W. and {Vecchiato}, A.},
    title = "{Gaia Data Release 2. The astrometric solution}",
  journal = {\aap},
archivePrefix = "arXiv",
   eprint = {1804.09366},
 primaryClass = "astro-ph.IM",
     year = 2018,
    month = aug,
   volume = 616,
      eid = {A2},
    pages = {A2},
      doi = {10.1051/0004-6361/201832727},
   adsurl = {http://adsabs.harvard.edu/abs/2018A%26A...616A...2L}
}

@ARTICLE{Burgasser2003,
       author = {{Burgasser}, Adam J. and {Kirkpatrick}, J. Davy and {Reid}, I. Neill and {Brown}, Michael E. and {Miskey}, Cherie L. and {Gizis}, John E.},
        title = "{Binarity in Brown Dwarfs: T Dwarf Binaries Discovered with the Hubble Space Telescope Wide Field Planetary Camera 2}",
      journal = {\apj},
         year = 2003,
        month = mar,
       volume = {586},
       number = {1},
        pages = {512-526},
          doi = {10.1086/346263},
archivePrefix = {arXiv},
       eprint = {astro-ph/0211470},
 primaryClass = {astro-ph},
       adsurl = {https://ui.adsabs.harvard.edu/abs/2003ApJ...586..512B}
}

@ARTICLE{Rosenthal2021,
       author = {{Rosenthal}, Lee J. and {Fulton}, Benjamin J. and {Hirsch}, Lea A. and {Isaacson}, Howard T. and {Howard}, Andrew W. and {Dedrick}, Cayla M. and {Sherstyuk}, Ilya A. and {Blunt}, Sarah C. and {Petigura}, Erik A. and {Knutson}, Heather A. and {Behmard}, Aida and {Chontos}, Ashley and {Crepp}, Justin R. and {Crossfield}, Ian J.~M. and {Dalba}, Paul A. and {Fischer}, Debra A. and {Henry}, Gregory W. and {Kane}, Stephen R. and {Kosiarek}, Molly and {Marcy}, Geoffrey W. and {Rubenzahl}, Ryan A. and {Weiss}, Lauren M. and {Wright}, Jason T.},
        title = "{The California Legacy Survey. I. A Catalog of 178 Planets from Precision Radial Velocity Monitoring of 719 Nearby Stars over Three Decades}",
      journal = {\apjs},
         year = 2021,
        month = jul,
       volume = {255},
       number = {1},
          eid = {8},
        pages = {8},
          doi = {10.3847/1538-4365/abe23c},
archivePrefix = {arXiv},
       eprint = {2105.11583},
 primaryClass = {astro-ph.EP},
       adsurl = {https://ui.adsabs.harvard.edu/abs/2021ApJS..255....8R}
}

@ARTICLE{Xiao2023,
       author = {{Xiao}, Guang-Yao and {Liu}, Yu-Juan and {Teng}, Huan-Yu and {Wang}, Wei and {Brandt}, Timothy D. and {Zhao}, Gang and {Zhao}, Fei and {Zhai}, Meng and {Gao}, Qi},
        title = "{The Masses of a Sample of Radial-velocity Exoplanets with Astrometric Measurements}",
      journal = {Research in Astronomy and Astrophysics},
         year = 2023,
        month = may,
       volume = {23},
       number = {5},
          eid = {055022},
        pages = {055022},
          doi = {10.1088/1674-4527/accb7e},
archivePrefix = {arXiv},
       eprint = {2303.12409},
 primaryClass = {astro-ph.EP},
       adsurl = {https://ui.adsabs.harvard.edu/abs/2023RAA....23e5022X}
}

@ARTICLE{Winn2022,
       author = {{Winn}, Joshua N.},
        title = "{Joint Constraints on Exoplanetary Orbits from Gaia DR3 and Doppler Data}",
      journal = {\aj},
         year = 2022,
        month = nov,
       volume = {164},
       number = {5},
          eid = {196},
        pages = {196},
          doi = {10.3847/1538-3881/ac9126},
archivePrefix = {arXiv},
       eprint = {2209.05516},
 primaryClass = {astro-ph.EP},
       adsurl = {https://ui.adsabs.harvard.edu/abs/2022AJ....164..196W}
}

@ARTICLE{Unger2023,
       author = {{Unger}, N. and {S{\'e}gransan}, D. and {Barbato}, D. and {Delisle}, J.-B. and {Sahlmann}, J. and {Holl}, B. and {Udry}, S.},
        title = "{Exploring the brown dwarf desert with precision radial velocities and Gaia DR3 astrometric orbits}",
      journal = {\aap},
         year = 2023,
        month = dec,
       volume = {680},
          eid = {A16},
        pages = {A16},
          doi = {10.1051/0004-6361/202347578},
archivePrefix = {arXiv},
       eprint = {2310.02758},
 primaryClass = {astro-ph.EP},
       adsurl = {https://ui.adsabs.harvard.edu/abs/2023A&A...680A..16U}
}

@ARTICLE{Holl2023,
       author = {{Holl}, B. and {Sozzetti}, A. and {Sahlmann}, J. and {Giacobbe}, P. and {S{\'e}gransan}, D. and {Unger}, N. and {Delisle}, J. -B. and {Barbato}, D. and {Lattanzi}, M.~G. and {Morbidelli}, R. and {Sosnowska}, D.},
        title = "{Gaia Data Release 3. Astrometric orbit determination with Markov chain Monte Carlo and genetic algorithms: Systems with stellar, sub-stellar, and planetary mass companions}",
      journal = {\aap},
         year = 2023,
        month = jun,
       volume = {674},
          eid = {A10},
        pages = {A10},
          doi = {10.1051/0004-6361/202244161},
archivePrefix = {arXiv},
       eprint = {2206.05439},
 primaryClass = {astro-ph.EP},
       adsurl = {https://ui.adsabs.harvard.edu/abs/2023A&A...674A..10H}
}

@ARTICLE{Perryman1997,
       author = {{Perryman}, M.~A.~C. and {Lindegren}, L. and {Kovalevsky}, J. and {Hoeg}, E. and {Bastian}, U. and {Bernacca}, P.~L. and {Cr{\'e}z{\'e}}, M. and {Donati}, F. and {Grenon}, M. and {Grewing}, M. and {van Leeuwen}, F. and {van der Marel}, H. and {Mignard}, F. and {Murray}, C.~A. and {Le Poole}, R.~S. and {Schrijver}, H. and {Turon}, C. and {Arenou}, F. and {Froeschl{\'e}}, M. and {Petersen}, C.~S.},
        title = "{The HIPPARCOS Catalogue}",
      journal = {\aap},
         year = 1997,
        month = jul,
       volume = {323},
        pages = {L49-L52},
       adsurl = {https://ui.adsabs.harvard.edu/abs/1997A&A...323L..49P}
}

@ARTICLE{vanLeeuwen2007,
       author = {{van Leeuwen}, F.},
        title = "{Validation of the new Hipparcos reduction}",
      journal = {\aap},
         year = 2007,
        month = nov,
       volume = {474},
       number = {2},
        pages = {653-664},
          doi = {10.1051/0004-6361:20078357},
archivePrefix = {arXiv},
       eprint = {0708.1752},
 primaryClass = {astro-ph},
       adsurl = {https://ui.adsabs.harvard.edu/abs/2007A&A...474..653V}
}

@ARTICLE{Halbwachs2023,
       author = {{Halbwachs}, Jean-Louis and {Pourbaix}, Dimitri and {Arenou}, Fr{\'e}d{\'e}ric and {Galluccio}, Laurent and {Guillout}, Patrick and {Bauchet}, Nathalie and {Marchal}, Olivier and {Sadowski}, Gilles and {Teyssier}, David},
        title = "{Gaia Data Release 3. Astrometric binary star processing}",
      journal = {\aap},
         year = 2023,
        month = jun,
       volume = {674},
          eid = {A9},
        pages = {A9},
          doi = {10.1051/0004-6361/202243969},
archivePrefix = {arXiv},
       eprint = {2206.05726},
 primaryClass = {astro-ph.SR},
       adsurl = {https://ui.adsabs.harvard.edu/abs/2023A&A...674A...9H}
}

@ARTICLE{Gan2023gaia,
       author = {{Gan}, Tianjun},
        title = "{Gaia Astrometry and MIKE+PFS Doppler Data Joint Analysis Reveals that HD 175167b is a Massive Cold Jupiter}",
      journal = {Research Notes of the American Astronomical Society},
         year = 2023,
        month = oct,
       volume = {7},
       number = {10},
          eid = {226},
        pages = {226},
          doi = {10.3847/2515-5172/ad0643},
archivePrefix = {arXiv},
       eprint = {2310.14568},
 primaryClass = {astro-ph.EP},
       adsurl = {https://ui.adsabs.harvard.edu/abs/2023RNAAS...7..226G}
}

@ARTICLE{Feng2022,
       author = {{Feng}, Fabo and {Butler}, R. Paul and {Vogt}, Steven S. and {Clement}, Matthew S. and {Tinney}, C.~G. and {Cui}, Kaiming and {Aizawa}, Masataka and {Jones}, Hugh R.~A. and {Bailey}, J. and {Burt}, Jennifer and {Carter}, B.~D. and {Crane}, Jeffrey D. and {Flammini Dotti}, Francesco and {Holden}, Bradford and {Ma}, Bo and {Ogihara}, Masahiro and {Oppenheimer}, Rebecca and {O'Toole}, S.~J. and {Shectman}, Stephen A. and {Wittenmyer}, Robert A. and {Wang}, Sharon X. and {Wright}, D.~J. and {Xuan}, Yifan},
        title = "{3D Selection of 167 Substellar Companions to Nearby Stars}",
      journal = {\apjs},
         year = 2022,
        month = sep,
       volume = {262},
       number = {1},
          eid = {21},
        pages = {21},
          doi = {10.3847/1538-4365/ac7e57},
archivePrefix = {arXiv},
       eprint = {2208.12720},
 primaryClass = {astro-ph.EP},
       adsurl = {https://ui.adsabs.harvard.edu/abs/2022ApJS..262...21F}
}

@ARTICLE{Ida2004II,
       author = {{Ida}, Shigeru and {Lin}, D.~N.~C.},
        title = "{Toward a Deterministic Model of Planetary Formation. II. The Formation and Retention of Gas Giant Planets around Stars with a Range of Metallicities}",
      journal = {\apj},
         year = 2004,
        month = nov,
       volume = {616},
       number = {1},
        pages = {567-572},
          doi = {10.1086/424830},
archivePrefix = {arXiv},
       eprint = {astro-ph/0408019},
 primaryClass = {astro-ph},
       adsurl = {https://ui.adsabs.harvard.edu/abs/2004ApJ...616..567I}
}

@ARTICLE{Marcy2000,
       author = {{Marcy}, Geoffrey W. and {Butler}, R. Paul},
        title = "{Planets Orbiting Other Suns}",
      journal = {\pasp},
         year = 2000,
        month = feb,
       volume = {112},
       number = {768},
        pages = {137-140},
          doi = {10.1086/316516},
       adsurl = {https://ui.adsabs.harvard.edu/abs/2000PASP..112..137M}
}
\bibliographystyle{aasjournal}



\end{document}